# Binary Black Hole Mergers in the First Advanced LIGO Observing Run

B. P. Abbott,[1] R. Abbott,[1] T. D. Abbott,[2] M. R. Abernathy,[3] F. Acernese,[4,5] K. Ackley,[6] C. Adams,[7] T. Adams,[8] P. Addesso,[9] R. X. Adhikari,[1] V. B. Adya,[10] C. Affeldt,[10] M. Agathos,[11] K. Agatsuma,[11] N. Aggarwal,[12] O. D. Aguiar,[13] L. Aiello,[14,15] A. Ain,[16] P. Ajith,[17] B. Allen,[10,18,19] A. Allocca,[20,21] P. A. Altin,[22] S. B. Anderson,[1] W. G. Anderson,[18] K. Arai,[1] M. C. Araya,[1] C. C. Arceneaux,[23] J. S. Areeda,[24] N. Arnaud,[25] K. G. Arun,[26] S. Ascenzi,[27,15] G. Ashton,[28] M. Ast,[29] S. M. Aston,[7] P. Astone,[30] P. Aufmuth,[19] C. Aulbert,[10] S. Babak,[31] P. Bacon,[32] M. K. M. Bader,[11] P. T. Baker,[33] F. Baldaccini,[34,35] G. Ballardin,[36] S. W. Ballmer,[37] J. C. Barayoga,[1] S. E. Barclay,[38] B. C. Barish,[1] D. Barker,[39] F. Barone,[4,5] B. Barr,[38] L. Barsotti,[12] M. Barsuglia,[32] D. Barta,[40] J. Bartlett,[39] I. Bartos,[41] R. Bassiri,[42] A. Basti,[20,21] J. C. Batch,[39] C. Baune,[10] V. Bavigadda,[36] M. Bazzan,[43,44] M. Bejger,[45] A. S. Bell,[38] B. K. Berger,[1] G. Bergmann,[10] C. P. L. Berry,[46] D. Bersanetti,[47,48] A. Bertolini,[11] J. Betzwieser,[7] S. Bhagwat,[37] R. Bhandare,[49] I. A. Bilenko,[50] G. Billingsley,[1] J. Birch,[7] R. Birney,[51] O. Birnholtz,[10] S. Biscans,[12] A. Bisht,[10,19] M. Bitossi,[36] C. Biwer,[37] M. A. Bizouard,[25] J. K. Blackburn,[1] C. D. Blair,[52] D. G. Blair,[52] R. M. Blair,[39] S. Bloemen,[53] O. Bock,[10] M. Boer,[54] G. Bogaert,[54] C. Bogan,[10] A. Bohe,[31] C. Bond,[46] F. Bondu,[55] R. Bonnand,[8] B. A. Boom,[11] R. Bork,[1] V. Boschi,[20,21] S. Bose,[56,16] Y. Bouffanais,[32] A. Bozzi,[36] C. Bradaschia,[21] P. R. Brady,[18] V. B. Braginsky,[50,†] M. Branchesi,[57,58] J. E. Brau,[59] T. Briant,[60] A. Brillet,[54] M. Brinkmann,[10] V. Brisson,[25] P. Brockill,[18] J. E. Broida,[61] A. F. Brooks,[1] D. A. Brown,[37] D. D. Brown,[46] N. M. Brown,[12] S. Brunett,[1] C. C. Buchanan,[2] A. Buikema,[12] T. Bulik,[62] H. J. Bulten,[63,11] A. Buonanno,[31,64] D. Buskulic,[8] C. Buy,[32] R. L. Byer,[42] M. Cabero,[10] L. Cadonati,[65] G. Cagnoli,[66,67] C. Cahillane,[1] J. Calderón Bustillo,[65] T. Callister,[1] E. Calloni,[68,5] J. B. Camp,[69] K. C. Cannon,[70] J. Cao,[71] C. D. Capano,[10] E. Capocasa,[32] F. Carbognani,[36] S. Caride,[72] J. Casanueva Diaz,[25] C. Casentini,[27,15] S. Caudill,[18] M. Cavaglià,[23] F. Cavalier,[25] R. Cavalieri,[36] G. Cella,[21] C. B. Cepeda,[1] L. Cerboni Baiardi,[57,58] G. Cerretani,[20,21] E. Cesarini,[27,15] S. J. Chamberlin,[73] M. Chan,[38] S. Chao,[74] P. Charlton,[75] E. Chassande-Mottin,[32] B. D. Cheeseboro,[76] H. Y. Chen,[77] Y. Chen,[78] C. Cheng,[74] A. Chincarini,[48] A. Chiummo,[36] H. S. Cho,[79] M. Cho,[64] J. H. Chow,[22] N. Christensen,[61] Q. Chu,[52] S. Chua,[60] S. Chung,[52] G. Ciani,[6] F. Clara,[39] J. A. Clark,[65] F. Cleva,[54] E. Coccia,[27,14] P.-F. Cohadon,[60] A. Colla,[80,30] C. G. Collette,[81] L. Cominsky,[82] M. Constancio, Jr.,[13] A. Conte,[80,30] L. Conti,[44] D. Cook,[39] T. R. Corbitt,[2] N. Cornish,[33] A. Corsi,[72] S. Cortese,[36] C. A. Costa,[13] M. W. Coughlin,[61] S. B. Coughlin,[83] J.-P. Coulon,[54] S. T. Countryman,[41] P. Couvares,[1] E. E. Cowan,[65] D. M. Coward,[52] M. J. Cowart,[7] D. C. Coyne,[1] R. Coyne,[72] K. Craig,[38] J. D. E. Creighton,[18] J. Cripe,[2] S. G. Crowder,[84] A. Cumming,[38] L. Cunningham,[38] E. Cuoco,[36] T. Dal Canton,[10] S. L. Danilishin,[38] S. D'Antonio,[15] K. Danzmann,[19,10] N. S. Darman,[85] A. Dasgupta,[86] C. F. Da Silva Costa,[6] V. Dattilo,[36] I. Dave,[49] M. Davier,[25] G. S. Davies,[38] E. J. Daw,[87] R. Day,[36] S. De,[37] D. DeBra,[42] G. Debreczeni,[40] J. Degallaix,[66] M. De Laurentis,[68,5] S. Deléglise,[60] W. Del Pozzo,[46] T. Denker,[10] T. Dent,[10] V. Dergachev,[1] R. De Rosa,[68,5] R. T. DeRosa,[7] R. DeSalvo,[9] R. C. Devine,[76] S. Dhurandhar,[16] M. C. Díaz,[88] L. Di Fiore,[5] M. Di Giovanni,[89,90] T. Di Girolamo,[68,5] A. Di Lieto,[20,21] S. Di Pace,[80,30] I. Di Palma,[31,80,30] A. Di Virgilio,[21] V. Dolique,[66] F. Donovan,[12] K. L. Dooley,[23] S. Doravari,[10] R. Douglas,[38] T. P. Downes,[18] M. Drago,[10] R. W. P. Drever,[1] J. C. Driggers,[39] M. Ducrot,[8] S. E. Dwyer,[39] T. B. Edo,[87] M. C. Edwards,[61] A. Effler,[7] H.-B. Eggenstein,[10] P. Ehrens,[1] J. Eichholz,[6,1] S. S. Eikenberry,[6] W. Engels,[78] R. C. Essick,[12] T. Etzel,[1] M. Evans,[12] T. M. Evans,[7] R. Everett,[73] M. Factourovich,[41] V. Fafone,[27,15] H. Fair,[37] S. Fairhurst,[91] X. Fan,[71] Q. Fang,[52] S. Farinon,[48] B. Farr,[77] W. M. Farr,[46] M. Favata,[92] M. Fays,[91] H. Fehrmann,[10] M. M. Fejer,[42] E. Fenyvesi,[93] I. Ferrante,[20,21] E. C. Ferreira,[13] F. Ferrini,[36] F. Fidecaro,[20,21] I. Fiori,[36] D. Fiorucci,[32] R. P. Fisher,[37] R. Flaminio,[66,94] M. Fletcher,[38] H. Fong,[95] J.-D. Fournier,[54] S. Frasca,[80,30] F. Frasconi,[21] Z. Frei,[93] A. Freise,[46] R. Frey,[59] V. Frey,[25] P. Fritschel,[12] V. V. Frolov,[7] P. Fulda,[6] M. Fyffe,[7] H. A. G. Gabbard,[23] S. Gaebel,[46] J. R. Gair,[96] L. Gammaitoni,[34] S. G. Gaonkar,[16] F. Garufi,[68,5] G. Gaur,[97,86] N. Gehrels,[69] G. Gemme,[48] P. Geng,[88] E. Genin,[36] A. Gennai,[21] J. George,[49] L. Gergely,[98] V. Germain,[8] Abhirup Ghosh,[17] Archisman Ghosh,[17] S. Ghosh,[53,11] J. A. Giaime,[2,7] K. D. Giardina,[7] A. Giazotto,[21] K. Gill,[99] A. Glaefke,[38] E. Goetz,[39] R. Goetz,[6] L. Gondan,[93] G. González,[2] J. M. Gonzalez Castro,[20,21] A. Gopakumar,[100] N. A. Gordon,[38] M. L. Gorodetsky,[50] S. E. Gossan,[1] M. Gosselin,[36] R. Gouaty,[8] A. Grado,[101,5] C. Graef,[38] P. B. Graff,[64] M. Granata,[66] A. Grant,[38] S. Gras,[12] C. Gray,[39] G. Greco,[57,58] A. C. Green,[46] P. Groot,[53] H. Grote,[10] S. Grunewald,[31] G. M. Guidi,[57,58] X. Guo,[71] A. Gupta,[16] M. K. Gupta,[86] K. E. Gushwa,[1] E. K. Gustafson,[1] R. Gustafson,[102] J. J. Hacker,[24] B. R. Hall,[56] E. D. Hall,[1] H. Hamilton,[103] G. Hammond,[38] M. Haney,[100] M. M. Hanke,[10] J. Hanks,[39] C. Hanna,[73] M. D. Hannam,[91] J. Hanson,[7] T. Hardwick,[2] J. Harms,[57,58] G. M. Harry,[3] I. W. Harry,[31] M. J. Hart,[38] M. T. Hartman,[6] C.-J. Haster,[46] K. Haughian,[38] J. Healy,[104] A. Heidmann,[60] M. C. Heintze,[7] H. Heitmann,[54] P. Hello,[25] G. Hemming,[36] M. Hendry,[38] I. S. Heng,[38] J. Hennig,[38] J. Henry,[104] A. W. Heptonstall,[1] M. Heurs,[10,19] S. Hild,[38] D. Hoak,[36] D. Hofman,[66] K. Holt,[7] D. E. Holz,[77] P. Hopkins,[91] J. Hough,[38] E. A. Houston,[38] E. J. Howell,[52] Y. M. Hu,[10] S. Huang,[74] E. A. Huerta,[105] D. Huet,[25] B. Hughey,[99] S. Husa,[106] S. H. Huttner,[38] T. Huynh-Dinh,[7] N. Indik,[10] D. R. Ingram,[39] R. Inta,[72] H. N. Isa,[38] J.-M. Isac,[60] M. Isi,[1] T. Isogai,[12] B. R. Iyer,[17] K. Izumi,[39] T. Jacqmin,[60] H. Jang,[79] K. Jani,[65] P. Jaranowski,[107] S. Jawahar,[108] L. Jian,[52] F. Jiménez-Forteza,[106] W. W. Johnson,[2] N. K. Johnson-McDaniel,[17] D. I. Jones,[28] R. Jones,[38] R. J. G. Jonker,[11] L. Ju,[52] Haris K,[109]

C. V. Kalaghatgi,[91] V. Kalogera,[83] S. Kandhasamy,[23] G. Kang,[79] J. B. Kanner,[1] S. J. Kapadia,[10] S. Karki,[59] K. S. Karvinen,[10] M. Kasprzack,[36,2] E. Katsavounidis,[12] W. Katzman,[7] S. Kaufer,[19] T. Kaur,[52] K. Kawabe,[39] F. Kéfélian,[54] M. S. Kehl,[95] D. Keitel,[106] D. B. Kelley,[37] W. Kells,[1] R. Kennedy,[87] J. S. Key,[88] F. Y. Khalili,[50] I. Khan,[14] S. Khan,[91] Z. Khan,[86] E. A. Khazanov,[110] N. Kijbunchoo,[39] Chi-Woong Kim,[79] Chunglee Kim,[79] J. Kim,[111] K. Kim,[112] N. Kim,[42] W. Kim,[113] Y.-M. Kim,[111] S. J. Kimbrell,[65] E. J. King,[113] P. J. King,[39] J. S. Kissel,[39] B. Klein,[83] L. Kleybolte,[29] S. Klimenko,[6] S. M. Koehlenbeck,[10] S. Koley,[11] V. Kondrashov,[1] A. Kontos,[12] M. Korobko,[29] W. Z. Korth,[1] I. Kowalska,[62] D. B. Kozak,[1] V. Kringel,[10] B. Krishnan,[10] A. Królak,[114,115] C. Krueger,[19] G. Kuehn,[10] P. Kumar,[95] R. Kumar,[86] L. Kuo,[74] A. Kutynia,[114] B. D. Lackey,[37] M. Landry,[39] J. Lange,[104] B. Lantz,[42] P. D. Lasky,[116] M. Laxen,[7] A. Lazzarini,[1] C. Lazzaro,[44] P. Leaci,[80,30] S. Leavey,[38] E. O. Lebigot,[32,71] C. H. Lee,[111] H. K. Lee,[112] H. M. Lee,[117] K. Lee,[38] A. Lenon,[37] M. Leonardi,[89,90] J. R. Leong,[10] N. Leroy,[25] N. Letendre,[8] Y. Levin,[116] J. B. Lewis,[1] T. G. F. Li,[118] A. Libson,[12] T. B. Littenberg,[119] N. A. Lockerbie,[108] A. L. Lombardi,[120] L. T. London,[91] J. E. Lord,[37] M. Lorenzini,[14,15] V. Loriette,[121] M. Lormand,[7] G. Losurdo,[58] J. D. Lough,[10,19] C. Lousto,[104] H. Lück,[19,10] A. P. Lundgren,[10] R. Lynch,[12] Y. Ma,[52] B. Machenschalk,[10] M. MacInnis,[12] D. M. Macleod,[2] F. Magaña-Sandoval,[37] L. Magaña Zertuche,[37] R. M. Magee,[56] E. Majorana,[30] I. Maksimovic,[121] V. Malvezzi,[27,15] N. Man,[54] I. Mandel,[46] V. Mandic,[84] V. Mangano,[38] G. L. Mansell,[22] M. Manske,[18] M. Mantovani,[36] F. Marchesoni,[122,35] F. Marion,[8] S. Márka,[41] Z. Márka,[41] A. S. Markosyan,[42] E. Maros,[1] F. Martelli,[57,58] L. Martellini,[54] I. W. Martin,[38] D. V. Martynov,[12] J. N. Marx,[1] K. Mason,[12] A. Masserot,[8] T. J. Massinger,[37] M. Masso-Reid,[38] S. Mastrogiovanni,[80,30] F. Matichard,[12] L. Matone,[41] N. Mavalvala,[12] N. Mazumder,[56] R. McCarthy,[39] D. E. McClelland,[22] S. McCormick,[7] S. C. McGuire,[123] G. McIntyre,[1] J. McIver,[1] D. J. McManus,[22] T. McRae,[22] S. T. McWilliams,[76] D. Meacher,[73] G. D. Meadors,[31,10] J. Meidam,[11] A. Melatos,[85] G. Mendell,[39] R. A. Mercer,[18] E. L. Merilh,[39] M. Merzougui,[54] S. Meshkov,[1] C. Messenger,[38] C. Messick,[73] R. Metzdorff,[60] P. M. Meyers,[84] F. Mezzani,[30,80] H. Miao,[46] C. Michel,[66] H. Middleton,[46] E. E. Mikhailov,[124] L. Milano,[68,5] A. L. Miller,[6,80,30] A. Miller,[83] B. B. Miller,[83] J. Miller,[12] M. Millhouse,[33] Y. Minenkov,[15] J. Ming,[31] S. Mirshekari,[125] C. Mishra,[17] S. Mitra,[16] V. P. Mitrofanov,[50] G. Mitselmakher,[6] R. Mittleman,[12] A. Moggi,[21] M. Mohan,[36] S. R. P. Mohapatra,[12] M. Montani,[57,58] B. C. Moore,[92] C. J. Moore,[126] D. Moraru,[39] G. Moreno,[39] S. R. Morriss,[88] K. Mossavi,[10] B. Mours,[8] C. M. Mow-Lowry,[46] G. Mueller,[6] A. W. Muir,[91] Arunava Mukherjee,[17] D. Mukherjee,[18] S. Mukherjee,[88] N. Mukund,[16] A. Mullavey,[7] J. Munch,[113] D. J. Murphy,[41] P. G. Murray,[38] A. Mytidis,[6] I. Nardecchia,[27,15] L. Naticchioni,[80,30] R. K. Nayak,[127] K. Nedkova,[120] G. Nelemans,[53,11] T. J. N. Nelson,[7] M. Neri,[47,48] A. Neunzert,[102] G. Newton,[38] T. T. Nguyen,[22] A. B. Nielsen,[10] S. Nissanke,[53,11] A. Nitz,[10] F. Nocera,[36] D. Nolting,[7] M. E. N. Normandin,[88] L. K. Nuttall,[37] J. Oberling,[39] E. Ochsner,[18] J. O'Dell,[128] E. Oelker,[12] G. H. Ogin,[129] J. J. Oh,[130] S. H. Oh,[130] F. Ohme,[91] M. Oliver,[106] P. Oppermann,[10] Richard J. Oram,[7] B. O'Reilly,[7] R. O'Shaughnessy,[104] D. J. Ottaway,[113] H. Overmier,[7] B. J. Owen,[72] A. Pai,[109] S. A. Pai,[49] J. R. Palamos,[59] O. Palashov,[110] C. Palomba,[30] A. Pal-Singh,[29] H. Pan,[74] Y. Pan,[64] C. Pankow,[83] F. Pannarale,[91] B. C. Pant,[49] F. Paoletti,[36,21] A. Paoli,[36] M. A. Papa,[31,18,10] H. R. Paris,[42] W. Parker,[7] D. Pascucci,[38] A. Pasqualetti,[36] R. Passaquieti,[20,21] D. Passuello,[21] B. Patricelli,[20,21] Z. Patrick,[42] B. L. Pearlstone,[38] M. Pedraza,[1] R. Pedurand,[66,131] L. Pekowsky,[37] A. Pele,[7] S. Penn,[132] A. Perreca,[1] L. M. Perri,[83] H. P. Pfeiffer,[95] M. Phelps,[38] O. J. Piccinni,[80,30] M. Pichot,[54] F. Piergiovanni,[57,58] V. Pierro,[9] G. Pillant,[36] L. Pinard,[66] I. M. Pinto,[9] M. Pitkin,[38] M. Poe,[18] R. Poggiani,[20,21] P. Popolizio,[36] E. Porter,[32] A. Post,[10] J. Powell,[38] J. Prasad,[16] V. Predoi,[91] T. Prestegard,[84] L. R. Price,[1] M. Prijatelj,[10,36] M. Principe,[9] S. Privitera,[31] R. Prix,[10] G. A. Prodi,[89,90] L. Prokhorov,[50] O. Puncken,[10] M. Punturo,[35] P. Puppo,[30] M. Pürrer,[31] H. Qi,[18] J. Qin,[52] S. Qiu,[116] V. Quetschke,[88] E. A. Quintero,[1] R. Quitzow-James,[59] F. J. Raab,[39] D. S. Rabeling,[22] H. Radkins,[39] P. Raffai,[93] S. Raja,[49] C. Rajan,[49] M. Rakhmanov,[88] P. Rapagnani,[80,30] V. Raymond,[31] M. Razzano,[20,21] V. Re,[27] J. Read,[24] C. M. Reed,[39] T. Regimbau,[54] L. Rei,[48] S. Reid,[51] D. H. Reitze,[1,6] H. Rew,[124] S. D. Reyes,[37] F. Ricci,[80,30] K. Riles,[102] M. Rizzo,[104] N. A. Robertson,[1,38] R. Robie,[38] F. Robinet,[25] A. Rocchi,[15] L. Rolland,[8] J. G. Rollins,[1] V. J. Roma,[59] J. D. Romano,[88] R. Romano,[4,5] G. Romanov,[124] J. H. Romie,[7] D. Rosińska,[133,45] S. Rowan,[38] A. Rüdiger,[10] P. Ruggi,[36] K. Ryan,[39] S. Sachdev,[1] T. Sadecki,[39] L. Sadeghian,[18] M. Sakellariadou,[134] L. Salconi,[36] M. Saleem,[109] F. Salemi,[10] A. Samajdar,[127] L. Sammut,[116] E. J. Sanchez,[1] V. Sandberg,[39] B. Sandeen,[83] J. R. Sanders,[37] B. Sassolas,[66] B. S. Sathyaprakash,[91] P. R. Saulson,[37] O. E. S. Sauter,[102] R. L. Savage,[39] A. Sawadsky,[19] P. Schale,[59] R. Schilling,[10,†] J. Schmidt,[10] P. Schmidt,[1,78] R. Schnabel,[29] R. M. S. Schofield,[59] A. Schönbeck,[29] E. Schreiber,[10] D. Schuette,[10,19] B. F. Schutz,[91,31] J. Scott,[38] S. M. Scott,[22] D. Sellers,[7] A. S. Sengupta,[97] D. Sentenac,[36] V. Sequino,[27,15] A. Sergeev,[110] Y. Setyawati,[53,11] D. A. Shaddock,[22] T. Shaffer,[39] M. S. Shahriar,[83] M. Shaltev,[10] B. Shapiro,[42] P. Shawhan,[64] A. Sheperd,[18] D. H. Shoemaker,[12] D. M. Shoemaker,[65] K. Siellez,[65] X. Siemens,[18] M. Sieniawska,[45] D. Sigg,[39] A. D. Silva,[13] A. Singer,[1] L. P. Singer,[69] A. Singh,[31,10,19] R. Singh,[2] A. Singhal,[14] A. M. Sintes,[106] B. J. J. Slagmolen,[22] J. R. Smith,[24] N. D. Smith,[1] R. J. E. Smith,[1] E. J. Son,[130] B. Sorazu,[38] F. Sorrentino,[48] T. Souradeep,[16] A. K. Srivastava,[86] A. Staley,[41] M. Steinke,[10] J. Steinlechner,[38] S. Steinlechner,[38] D. Steinmeyer,[10,19] B. C. Stephens,[18] S. Stevenson,[46] R. Stone,[88] K. A. Strain,[38] N. Straniero,[66] G. Stratta,[57,58] N. A. Strauss,[61] S. Strigin,[50] R. Sturani,[125] A. L. Stuver,[7] T. Z. Summerscales,[135] L. Sun,[85] S. Sunil,[86] P. J. Sutton,[91] B. L. Swinkels,[36] M. J. Szczepańczyk,[99] M. Tacca,[32] D. Talukder,[59] D. B. Tanner,[6] M. Tápai,[98] S. P. Tarabrin,[10] A. Taracchini,[31] R. Taylor,[1] T. Theeg,[10] M. P. Thirugnanasambandam,[1] E. G. Thomas,[46] M. Thomas,[7]

P. Thomas,[39] K. A. Thorne,[7] E. Thrane,[116] S. Tiwari,[14,90] V. Tiwari,[91] K. V. Tokmakov,[108] K. Toland,[38] C. Tomlinson,[87] M. Tonelli,[20,21] Z. Tornasi,[38] C. V. Torres,[88,†] C. I. Torrie,[1] D. Töyrä,[46] F. Travasso,[34,35] G. Traylor,[7] D. Trifirò,[23] M. C. Tringali,[89,90] L. Trozzo,[136,21] M. Tse,[12] M. Turconi,[54] D. Tuyenbayev,[88] D. Ugolini,[137] C. S. Unnikrishnan,[100] A. L. Urban,[18] S. A. Usman,[37] H. Vahlbruch,[19] G. Vajente,[1] G. Valdes,[88] M. Vallisneri,[78] N. van Bakel,[11] M. van Beuzekom,[11] J. F. J. van den Brand,[63,11] C. Van Den Broeck,[11] D. C. Vander-Hyde,[37] L. van der Schaaf,[11] J. V. van Heijningen,[11] A. A. van Veggel,[38] M. Vardaro,[43,44] S. Vass,[1] M. Vasúth,[40] R. Vaulin,[12] A. Vecchio,[46] G. Vedovato,[44] J. Veitch,[46] P. J. Veitch,[113] K. Venkateswara,[138] D. Verkindt,[8] F. Vetrano,[57,58] A. Viceré,[57,58] S. Vinciguerra,[46] D. J. Vine,[51] J.-Y. Vinet,[54] S. Vitale,[12] T. Vo,[37] H. Vocca,[34,35] C. Vorvick,[39] D. V. Voss,[6] W. D. Vousden,[46] S. P. Vyatchanin,[50] A. R. Wade,[22] L. E. Wade,[139] M. Wade,[139] M. Walker,[2] L. Wallace,[1] S. Walsh,[31,10] G. Wang,[14,58] H. Wang,[46] M. Wang,[46] X. Wang,[71] Y. Wang,[52] R. L. Ward,[22] J. Warner,[39] M. Was,[8] B. Weaver,[39] L.-W. Wei,[54] M. Weinert,[10] A. J. Weinstein,[1] R. Weiss,[12] L. Wen,[52] P. Weßels,[10] T. Westphal,[10] K. Wette,[10] J. T. Whelan,[104] S. E. Whitcomb,[1] B. F. Whiting,[6] R. D. Williams,[1] A. R. Williamson,[91] J. L. Willis,[103] B. Willke,[19,10] M. H. Wimmer,[10,19] W. Winkler,[10] C. C. Wipf,[1] H. Wittel,[10,19] G. Woan,[38] J. Woehler,[10] J. Worden,[39] J. L. Wright,[38] D. S. Wu,[10] G. Wu,[7] J. Yablon,[83] W. Yam,[12] H. Yamamoto,[1] C. C. Yancey,[64] H. Yu,[12] M. Yvert,[8] A. Zadrożny,[114] L. Zangrando,[44] M. Zanolin,[99] J.-P. Zendri,[44] M. Zevin,[83] L. Zhang,[1] M. Zhang,[124] Y. Zhang,[104] C. Zhao,[52] M. Zhou,[83] Z. Zhou,[83] X. J. Zhu,[52] M. E. Zucker,[1,12] S. E. Zuraw,[120] and J. Zweizig[1]

(LIGO Scientific Collaboration and Virgo Collaboration)[*]

[1]*LIGO, California Institute of Technology, Pasadena, California 91125, USA*
[2]*Louisiana State University, Baton Rouge, Louisiana 70803, USA*
[3]*American University, Washington, D.C. 20016, USA*
[4]*Università di Salerno, Fisciano, I-84084 Salerno, Italy*
[5]*INFN, Sezione di Napoli, Complesso Universitario di Monte S.Angelo, I-80126 Napoli, Italy*
[6]*University of Florida, Gainesville, Florida 32611, USA*
[7]*LIGO Livingston Observatory, Livingston, Louisiana 70754, USA*
[8]*Laboratoire d'Annecy-le-Vieux de Physique des Particules (LAPP), Université Savoie Mont Blanc, CNRS/IN2P3, F-74941 Annecy-le-Vieux, France*
[9]*University of Sannio at Benevento, I-82100 Benevento, Italy and INFN, Sezione di Napoli, I-80100 Napoli, Italy*
[10]*Albert-Einstein-Institut, Max-Planck-Institut für Gravitationsphysik, D-30167 Hannover, Germany*
[11]*Nikhef, Science Park, 1098 XG Amsterdam, The Netherlands*
[12]*LIGO, Massachusetts Institute of Technology, Cambridge, Massachusetts 02139, USA*
[13]*Instituto Nacional de Pesquisas Espaciais, 12227-010 São José dos Campos, São Paulo, Brazil*
[14]*INFN, Gran Sasso Science Institute, I-67100L'Aquila, Italy*
[15]*INFN, Sezione di Roma Tor Vergata, I-00133 Roma, Italy*
[16]*Inter-University Centre for Astronomy and Astrophysics, Pune 411007, India*
[17]*International Centre for Theoretical Sciences, Tata Institute of Fundamental Research, Bangalore 560012, India*
[18]*University of Wisconsin-Milwaukee, Milwaukee, Wisconsin 53201, USA*
[19]*Leibniz Universität Hannover, D-30167 Hannover, Germany*
[20]*Università di Pisa, I-56127 Pisa, Italy*
[21]*INFN, Sezione di Pisa, I-56127 Pisa, Italy*
[22]*Australian National University, Canberra, Australian Capital Territory 0200, Australia*
[23]*The University of Mississippi, University, Mississippi 38677, USA*
[24]*California State University Fullerton, Fullerton, California 92831, USA*
[25]*LAL, Université Paris-Sud, CNRS/IN2P3, Université Paris-Saclay, F-91898 Orsay, France*
[26]*Chennai Mathematical Institute, Chennai 603103, India*
[27]*Università di Roma Tor Vergata, I-00133 Roma, Italy*
[28]*University of Southampton, Southampton SO17 1BJ, United Kingdom*
[29]*Universität Hamburg, D-22761 Hamburg, Germany*
[30]*INFN, Sezione di Roma, I-00185 Roma, Italy*
[31]*Albert-Einstein-Institut, Max-Planck-Institut für Gravitations-physik, D-14476 Potsdam-Golm, Germany*
[32]*APC, AstroParticule et Cosmologie, Université Paris Diderot, CNRS/IN2P3, CEA/Irfu, Observatoire de Paris, Sorbonne Paris Cité, F-75205 Paris Cedex 13, France*
[33]*Montana State University, Bozeman, Montana 59717, USA*
[34]*Università di Perugia, I-06123 Perugia, Italy*
[35]*INFN, Sezione di Perugia, I-06123 Perugia, Italy*
[36]*European Gravitational Observatory (EGO), I-56021 Cascina, Pisa, Italy*

[37]*Syracuse University, Syracuse, New York 13244, USA*
[38]*SUPA, University of Glasgow, Glasgow G12 8QQ, United Kingdom*
[39]*LIGO Hanford Observatory, Richland, Washington 99352, USA*
[40]*Wigner RCP, RMKI, H-1121 Budapest, Konkoly Thege Miklós út 29-33, Hungary*
[41]*Columbia University, New York, New York 10027, USA*
[42]*Stanford University, Stanford, California 94305, USA*
[43]*Università di Padova, Dipartimento di Fisica e Astronomia, I-35131 Padova, Italy*
[44]*INFN, Sezione di Padova, I-35131 Padova, Italy*
[45]*CAMK-PAN, 00-716 Warsaw, Poland*
[46]*University of Birmingham, Birmingham B15 2TT, United Kingdom*
[47]*Università degli Studi di Genova, I-16146 Genova, Italy*
[48]*INFN, Sezione di Genova, I-16146 Genova, Italy*
[49]*RRCAT, Indore, Madhya Pradesh 452013, India*
[50]*Faculty of Physics, Lomonosov Moscow State University, Moscow 119991, Russia*
[51]*SUPA, University of the West of Scotland, Paisley PA1 2BE, United Kingdom*
[52]*University of Western Australia, Crawley, Western Australia 6009, Australia*
[53]*Department of Astrophysics/IMAPP, Radboud University Nijmegen, P.O. Box 9010, 6500 GL Nijmegen, The Netherlands*
[54]*Artemis, Université Côte d'Azur, CNRS, Observatoire Côte d'Azur, CS 34229, F-06304 Nice cedex 4, France*
[55]*Institut de Physique de Rennes, CNRS, Université de Rennes 1, F-35042 Rennes, France*
[56]*Washington State University, Pullman, Washington 99164, USA*
[57]*Università degli Studi di Urbino "Carlo Bo", I-61029 Urbino, Italy*
[58]*INFN, Sezione di Firenze, I-50019 Sesto Fiorentino, Firenze, Italy*
[59]*University of Oregon, Eugene, Oregon 97403, USA*
[60]*Laboratoire Kastler Brossel, UPMC-Sorbonne Universités, CNRS, ENS-PSL Research University, Collège de France, F-75005 Paris, France*
[61]*Carleton College, Northfield, Minnesota 55057, USA*
[62]*Astronomical Observatory Warsaw University, 00-478 Warsaw, Poland*
[63]*VU University Amsterdam, 1081 HV Amsterdam, The Netherlands*
[64]*University of Maryland, College Park, Maryland 20742, USA*
[65]*Center for Relativistic Astrophysics and School of Physics, Georgia Institute of Technology, Atlanta, Georgia 30332, USA*
[66]*Laboratoire des Matériaux Avancés (LMA), CNRS/IN2P3, F-69622 Villeurbanne, France*
[67]*Université Claude Bernard Lyon 1, F-69622 Villeurbanne, France*
[68]*Università di Napoli "Federico II", Complesso Universitario di Monte S.Angelo, I-80126 Napoli, Italy*
[69]*NASA/Goddard Space Flight Center, Greenbelt, Maryland 20771, USA*
[70]*RESCEU, University of Tokyo, Tokyo 113-0033, Japan*
[71]*Tsinghua University, Beijing 100084, China*
[72]*Texas Tech University, Lubbock, Texas 79409, USA*
[73]*The Pennsylvania State University, University Park, Pennsylvania 16802, USA*
[74]*National Tsing Hua University, Hsinchu City, 30013 Taiwan, Republic of China*
[75]*Charles Sturt University, Wagga Wagga, New South Wales 2678, Australia*
[76]*West Virginia University, Morgantown, West Virginia 26506, USA*
[77]*University of Chicago, Chicago, Illinois 60637, USA*
[78]*Caltech CaRT, Pasadena, California 91125, USA*
[79]*Korea Institute of Science and Technology Information, Daejeon 305-806, Korea*
[80]*Università di Roma "La Sapienza", I-00185 Roma, Italy*
[81]*University of Brussels, Brussels 1050, Belgium*
[82]*Sonoma State University, Rohnert Park, California 94928, USA*
[83]*Center for Interdisciplinary Exploration & Research in Astrophysics (CIERA), Northwestern University, Evanston, Illinois 60208, USA*
[84]*University of Minnesota, Minneapolis, Minnesota 55455, USA*
[85]*The University of Melbourne, Parkville, Victoria 3010, Australia*
[86]*Institute for Plasma Research, Bhat, Gandhinagar 382428, India*
[87]*The University of Sheffield, Sheffield S10 2TN, United Kingdom*
[88]*The University of Texas Rio Grande Valley, Brownsville, Texas 78520, USA*
[89]*Università di Trento, Dipartimento di Fisica, I-38123 Povo, Trento, Italy*
[90]*INFN, Trento Institute for Fundamental Physics and Applications, I-38123 Povo, Trento, Italy*
[91]*Cardiff University, Cardiff CF24 3AA, United Kingdom*
[92]*Montclair State University, Montclair, New Jersey 07043, USA*

[93]*MTA Eötvös University, "Lendulet" Astrophysics Research Group, Budapest 1117, Hungary*
[94]*National Astronomical Observatory of Japan, 2-21-1 Osawa, Mitaka, Tokyo 181-8588, Japan*
[95]*Canadian Institute for Theoretical Astrophysics, University of Toronto, Toronto, Ontario M5S 3H8, Canada*
[96]*School of Mathematics, University of Edinburgh, Edinburgh EH9 3FD, United Kingdom*
[97]*Indian Institute of Technology, Gandhinagar, Ahmedabad, Gujarat 382424, India*
[98]*University of Szeged, Dóm tér 9, Szeged 6720, Hungary*
[99]*Embry-Riddle Aeronautical University, Prescott, Arizona 86301, USA*
[100]*Tata Institute of Fundamental Research, Mumbai 400005, India*
[101]*INAF, Osservatorio Astronomico di Capodimonte, I-80131 Napoli, Italy*
[102]*University of Michigan, Ann Arbor, Michigan 48109, USA*
[103]*Abilene Christian University, Abilene, Texas 79699, USA*
[104]*Rochester Institute of Technology, Rochester, New York 14623, USA*
[105]*NCSA, University of Illinois at Urbana-Champaign, Urbana, Illinois 61801, USA*
[106]*Universitat de les Illes Balears, IAC3—IEEC, E-07122 Palma de Mallorca, Spain*
[107]*University of Białystok, 15-424 Białystok, Poland*
[108]*SUPA, University of Strathclyde, Glasgow G1 1XQ, United Kingdom*
[109]*IISER-TVM, CET Campus, Trivandrum Kerala 695016, India*
[110]*Institute of Applied Physics, Nizhny Novgorod 603950, Russia*
[111]*Pusan National University, Busan 609-735, Korea*
[112]*Hanyang University, Seoul 133-791, Korea*
[113]*University of Adelaide, Adelaide, South Australia 5005, Australia*
[114]*NCBJ, 05-400 Świerk-Otwock, Poland*
[115]*IM-PAN, 00-956 Warsaw, Poland*
[116]*Monash University, Victoria 3800, Australia*
[117]*Seoul National University, Seoul 151-742, Korea*
[118]*The Chinese University of Hong Kong, Shatin, NT, Hong Kong SAR, China*
[119]*University of Alabama in Huntsville, Huntsville, Alabama 35899, USA*
[120]*University of Massachusetts-Amherst, Amherst, Massachusetts 01003, USA*
[121]*ESPCI, CNRS, F-75005 Paris, France*
[122]*Università di Camerino, Dipartimento di Fisica, I-62032 Camerino, Italy*
[123]*Southern University and A&M College, Baton Rouge, Louisiana 70813, USA*
[124]*College of William and Mary, Williamsburg, Virginia 23187, USA*
[125]*Instituto de Física Teórica, University Estadual Paulista/ICTP South American Institute for Fundamental Research, São Paulo, São Paulo 01140-070, Brazil*
[126]*University of Cambridge, Cambridge CB2 1TN, United Kingdom*
[127]*IISER-Kolkata, Mohanpur, West Bengal 741252, India*
[128]*Rutherford Appleton Laboratory, HSIC, Chilton, Didcot, Oxon OX11 0QX, United Kingdom*
[129]*Whitman College, 345 Boyer Avenue, Walla Walla, Washington 99362, USA*
[130]*National Institute for Mathematical Sciences, Daejeon 305-390, Korea*
[131]*Université de Lyon, F-69361 Lyon, France*
[132]*Hobart and William Smith Colleges, Geneva, New York 14456, USA*
[133]*Janusz Gil Institute of Astronomy, University of Zielona Góra, 65-265 Zielona Góra, Poland*
[134]*King's College London, University of London, London WC2R 2LS, United Kingdom*
[135]*Andrews University, Berrien Springs, Michigan 49104, USA*
[136]*Università di Siena, I-53100 Siena, Italy*
[137]*Trinity University, San Antonio, Texas 78212, USA*
[138]*University of Washington, Seattle, Washington 98195, USA*
[139]*Kenyon College, Gambier, Ohio 43022, USA*

The first observational run of the Advanced LIGO detectors, from September 12, 2015 to January 19, 2016, saw the first detections of gravitational waves from binary black hole mergers. In this paper, we present full results from a search for binary black hole merger signals with total masses up to $100 M_\odot$ and detailed implications from our observations of these systems. Our search, based on general-relativistic models of gravitational-wave signals from binary black hole systems, unambiguously identified two signals, GW150914 and GW151226, with a significance of greater than $5\sigma$ over the observing period. It also identified a third possible signal, LVT151012, with substantially lower significance and with an 87% probability of being of astrophysical origin. We provide detailed estimates of the parameters of the observed systems. Both GW150914 and GW151226 provide an unprecedented opportunity to study the two-body motion of a compact-object binary in the large velocity, highly nonlinear regime. We do not observe any deviations from general relativity, and we place improved empirical bounds on several high-order post-Newtonian coefficients. From our observations, we infer stellar-mass binary black hole merger rates lying in the range 9–240 $\mathrm{Gpc}^{-3}\,\mathrm{yr}^{-1}$. These observations are beginning to inform astrophysical predictions of binary black hole formation rates and indicate that future observing runs of the Advanced detector network will yield many more gravitational-wave detections.

[*]Full author list given at the end of the article.
[†]Deceased.

## I. INTRODUCTION

The first observing run (O1) of the Advanced LIGO detectors took place from September 12, 2015, to January 19, 2016. The detectors provided unprecedented sensitivity to gravitational waves over a range of frequencies from 30 Hz to several kHz [1], which covers the frequencies of gravitational waves emitted during the late inspiral, merger, and ringdown of stellar-mass binary black holes (BBHs). In this paper, we report the results of a matched-filter search using relativistic models of BBH waveforms during the whole of the first Advanced LIGO observing run. The compact binary coalescence (CBC) search targets gravitational-wave emission from compact-object binaries with individual masses from $1\mathrm{M}_\odot$ to $99\mathrm{M}_\odot$, total mass less than $100\mathrm{M}_\odot$, and dimensionless spins up to 0.99. Here, we report on results of the search for BBHs. The search was performed using two independently implemented analyses, referred to as PyCBC [2–4] and GstLAL [5–7]. These analyses use a common set of template waveforms [8–10] but differ in their implementations of matched filtering [11,12], their use of detector data-quality information [13], the techniques used to mitigate the effect of non-Gaussian noise transients in the detector [5,14], and the methods for estimating the noise background of the search [3,15]. We obtain results that are consistent between the two analyses.

The search identified two BBH mergers: GW150914, observed on September 14, 2015 at 09:50:45 UTC [16], and GW151226, observed on December 26, 2015 at 03:38:53 UTC [17]. Both of these signals were observed with a significance greater than $5\sigma$. In addition, a third candidate event, LVT151012, consistent with a BBH merger was observed on October 12, 2015 at 09:54:43 UTC with a significance of $\lesssim 2\sigma$. Although LVT151012 is not significant enough to claim an unambiguous detection, it is more likely to have resulted from a gravitational-wave signal than from an instrumental or environmental noise transient. The key parameters of the events are summarized in Table I.

The properties of the sources can be inferred from the observed gravitational waveforms. In particular, the binary evolution, which is encoded in the phasing of the gravitational-wave signal, is governed by the masses and spins of the binary components. The sky location of the source is primarily determined through time of arrival differences at the two Advanced LIGO sites. The observed amplitudes and relative phase of the signal in the two Advanced LIGO detectors can be used to further restrict the sky location and infer the distance to the source and the binary orientation. We provide a detailed evaluation of the source properties and inferred parameters of GW150914, GW151226, and LVT151012. We use models of the waveform covering the inspiral, merger, and ringdown phases based on combining post-Newtonian (PN) theory [19–24], the effective-one-body (EOB) formalism [25–29], and numerical relativity simulations [30–36]. One model is restricted to spins aligned with the orbital angular momentum [8,9], while the other allows for nonaligned orientation of the spins, which can lead to precession of the orbital plane [37,38]. The parameters of GW150914 have been reported previously in Refs. [39,40]. We provide revised results which make use of updated instrumental calibration.

The emitted signals depend upon the strong field dynamics of general relativity; thus, our observations provide an extraordinary opportunity to test the predictions of general relativity for binary coalescence waveforms. Several tests of general relativity were performed using GW150914, as described in Ref. [41]. One of these was a parametrized test for the consistency of the observed waveform with a general-relativity-based model. We perform a similar test on GW151226. Since this source is of lower mass than GW150914, the observed waveform lasts for many more cycles in the detector data, allowing us to better constrain the PN coefficients that describe the evolution of the binary through the inspiral phase. In addition, we combine the results from GW150914 and GW151226 to place still tighter bounds on deviations from general relativity.

The observed events begin to reveal a population of stellar-mass black hole mergers. We use these signals to constrain the rates of BBH mergers in the universe and begin to probe the mass distribution of black hole mergers. The inferred rates are consistent with those derived from GW150914 [42]. We also discuss the astrophysical implications of the observations and the prospects for future Advanced LIGO and Virgo observing runs.

TABLE I. Details of the three most significant events. The false alarm rate, p-value, and significance are from the PyCBC analysis; the GstLAL results are consistent with this. For source parameters, we report median values with 90% credible intervals that include statistical errors, and systematic errors from averaging the results of different waveform models. The uncertainty for the peak luminosity includes an estimate of additional error from the fitting formula. The sky localization is the area of the 90% credible area. Masses are given in the source frame; to convert to the detector frame, multiply by $(1+z)$. The source redshift assumes standard cosmology [18].

| Event | GW150914 | GW151226 | LVT151012 |
|---|---|---|---|
| Signal-to-noise ratio $\rho$ | 23.7 | 13.0 | 9.7 |
| False alarm rate FAR/yr$^{-1}$ | $< 6.0 \times 10^{-7}$ | $< 6.0 \times 10^{-7}$ | 0.37 |
| p-value | $7.5 \times 10^{-8}$ | $7.5 \times 10^{-8}$ | 0.045 |
| Significance | $> 5.3\sigma$ | $> 5.3\sigma$ | $1.7\sigma$ |
| Primary mass $m_1^{\text{source}}/\text{M}_\odot$ | $36.2^{+5.2}_{-3.8}$ | $14.2^{+8.3}_{-3.7}$ | $23^{+18}_{-6}$ |
| Secondary mass $m_2^{\text{source}}/\text{M}_\odot$ | $29.1^{+3.7}_{-4.4}$ | $7.5^{+2.3}_{-2.3}$ | $13^{+4}_{-5}$ |
| Chirp mass $\mathcal{M}^{\text{source}}/\text{M}_\odot$ | $28.1^{+1.8}_{-1.5}$ | $8.9^{+0.3}_{-0.3}$ | $15.1^{+1.4}_{-1.1}$ |
| Total mass $M^{\text{source}}/\text{M}_\odot$ | $65.3^{+4.1}_{-3.4}$ | $21.8^{+5.9}_{-1.7}$ | $37^{+13}_{-4}$ |
| Effective inspiral spin $\chi_{\text{eff}}$ | $-0.06^{+0.14}_{-0.14}$ | $0.21^{+0.20}_{-0.10}$ | $0.0^{+0.3}_{-0.2}$ |
| Final mass $M_{\text{f}}^{\text{source}}/\text{M}_\odot$ | $62.3^{+3.7}_{-3.1}$ | $20.8^{+6.1}_{-1.7}$ | $35^{+14}_{-4}$ |
| Final spin $a_{\text{f}}$ | $0.68^{+0.05}_{-0.06}$ | $0.74^{+0.06}_{-0.06}$ | $0.66^{+0.09}_{-0.10}$ |
| Radiated energy $E_{\text{rad}}/(\text{M}_\odot c^2)$ | $3.0^{+0.5}_{-0.4}$ | $1.0^{+0.1}_{-0.2}$ | $1.5^{+0.3}_{-0.4}$ |
| Peak luminosity $\ell_{\text{peak}}/(\text{erg s}^{-1})$ | $3.6^{+0.5}_{-0.4} \times 10^{56}$ | $3.3^{+0.8}_{-1.6} \times 10^{56}$ | $3.1^{+0.8}_{-1.8} \times 10^{56}$ |
| Luminosity distance $D_{\text{L}}/\text{Mpc}$ | $420^{+150}_{-180}$ | $440^{+180}_{-190}$ | $1000^{+500}_{-500}$ |
| Source redshift $z$ | $0.09^{+0.03}_{-0.04}$ | $0.09^{+0.03}_{-0.04}$ | $0.20^{+0.09}_{-0.09}$ |
| Sky localization $\Delta\Omega/\text{deg}^2$ | 230 | 850 | 1600 |

The results presented here are restricted to BBH systems with total masses less than $100\text{M}_\odot$. Searches for compact binary systems containing neutron stars are presented in Ref. [43], and searches for more massive black holes and unmodeled transient signals will be reported elsewhere.

This paper is organized as follows: Section II provides an overview of the Advanced LIGO detectors during the first observing run, as well as the data used in the search. Section III presents the results of the search, details of the two gravitational-wave events, GW150914 and GW151226, and the candidate event LVT151012. Section IV provides detailed parameter-estimation results for the events. Section V presents results for the consistency of the two events, GW150914 and GW151226, with the predictions of general relativity. Section VI presents the inferred rate of stellar-mass BBH mergers, and Sec. VII discusses the implications of these observations and future prospects. We include appendixes that provide additional technical details of the methods used. Appendix A describes the CBC search, with A 1 and A 2 presenting details of the construction and tuning of the two independently implemented analyses used in the search, highlighting differences from the methods described in Ref. [44]. Appendix B provides a description of the parameter-estimation analysis and includes a summary table of results for all three events. Appendixes C and D provide details of the methods used to infer merger rates and mass distributions, respectively.

## II. OVERVIEW OF THE INSTRUMENTS AND DATA SET

The two Advanced LIGO detectors, one located in Hanford, Washington (H1) and one in Livingston, Louisiana (L1), are modified Michelson interferometers with 4-km-long arms. The interferometer mirrors act as test masses, and the passage of a gravitational wave induces a differential arm length change which is proportional to the gravitational-wave strain amplitude. The Advanced LIGO detectors came online in September 2015 after a major upgrade targeting a tenfold improvement in sensitivity over the initial LIGO detectors [45]. While not yet operating at design sensitivity, both detectors reached an instrument noise 3–4 times lower than ever measured before in their most sensitive frequency band between 100 Hz and 300 Hz [1]. The corresponding observable volume of space for BBH mergers, in the mass range reported in this paper, was about 30 times greater, enabling the successful search reported here.

The typical instrument noise of the Advanced LIGO detectors during O1 is described in detail in Ref. [46]. In the left panel of Fig. 1, we show the amplitude spectral density of the total strain noise of both detectors, $\sqrt{S(f)}$, calibrated in units of strain per $\sqrt{\text{Hz}}$ [47]. Overlaid on the noise curves of the detectors, the waveforms of GW150914, GW151226, and LVT151012 are also shown. The expected signal-to-noise ratio (SNR) $\rho$ of a signal, $h(t)$, can be expressed as

$$\rho^2 = \int_0^\infty \frac{(2|\tilde{h}(f)|\sqrt{f})^2}{S_n(f)} \mathrm{d}\ln(f), \qquad (1)$$

where $\tilde{h}(f)$ is the Fourier transform of the signal. Writing it in this form motivates the normalization of the waveform plotted in Fig. 1, as the area between the signal and noise curves is indicative of the SNR of the events.

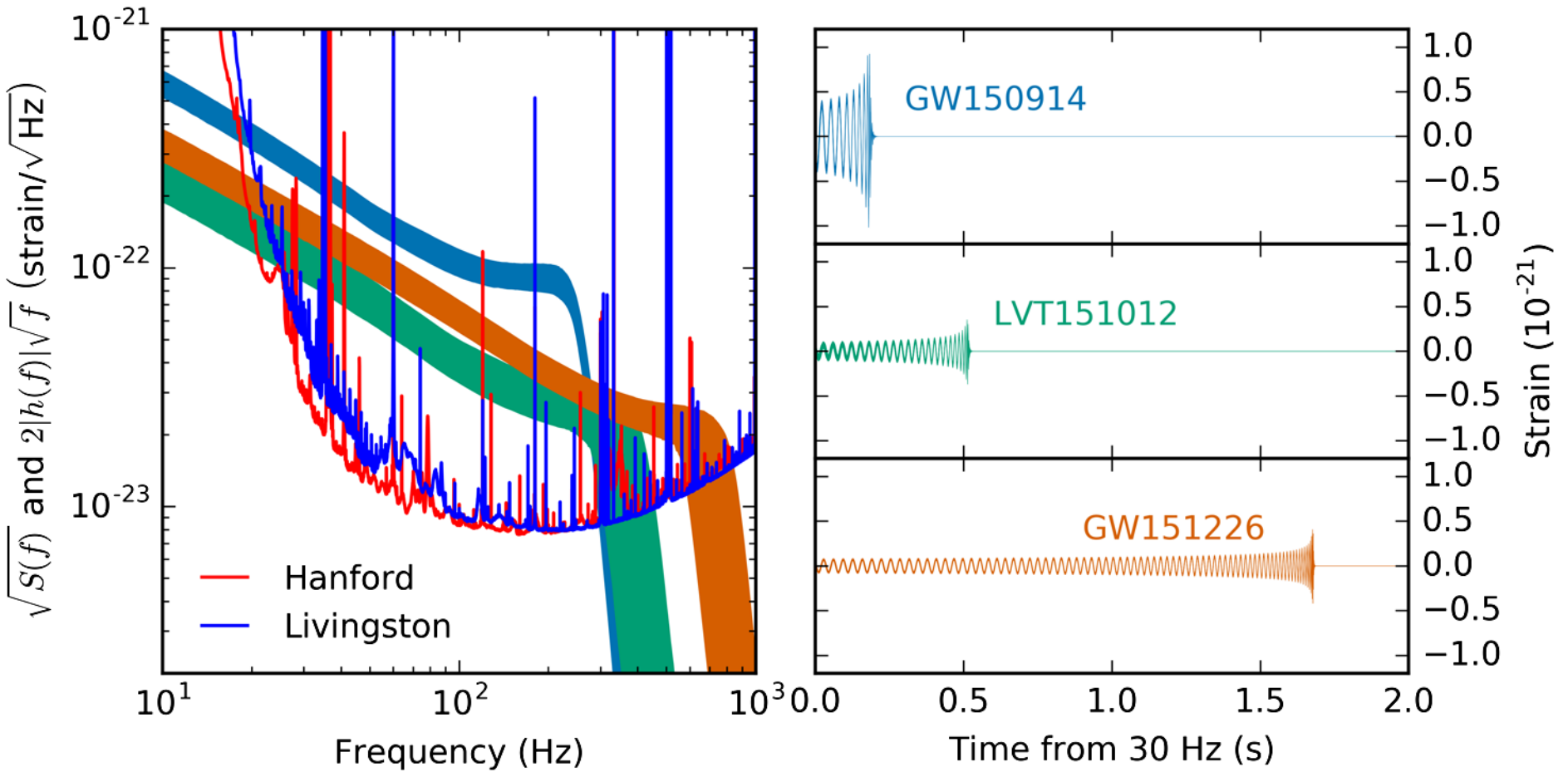

FIG. 1. Left panel: Amplitude spectral density of the total strain noise of the H1 and L1 detectors, $\sqrt{S(f)}$, in units of strain per $\sqrt{\mathrm{Hz}}$, and the recovered signals of GW150914, GW151226, and LVT151012 plotted so that the relative amplitudes can be related to the SNR of the signal (as described in the text). Right panel: Time evolution of the recovered signals from when they enter the detectors' sensitive band at 30 Hz. Both figures show the 90% credible regions of the LIGO Hanford signal reconstructions from a coherent Bayesian analysis using a nonprecessing spin waveform model [48].

The gravitational-wave signal from a BBH merger takes the form of a chirp, increasing in frequency and amplitude as the black holes spiral inwards. The amplitude of the signal is maximum at the merger, after which it decays rapidly as the final black hole rings down to equilibrium. In the frequency domain, the amplitude decreases with frequency during inspiral, as the signal spends a greater number of cycles at lower frequencies. This is followed by a slower falloff during merger and then a steep decrease during the ringdown. The amplitude of GW150914 is significantly larger than the other two events, and at the time of the merger, the gravitational-wave signal lies well above the noise. GW151226 has a lower amplitude but sweeps across the whole detector's sensitive band up to nearly 800 Hz. The corresponding time series of the three waveforms are plotted in the right panel of Fig. 1 to better visualize the difference in duration within the Advanced LIGO band: GW150914 lasts only a few cycles, while LVT151012 and GW151226 have lower amplitudes but last longer.

The analysis presented in this paper includes the total set of O1 data from September 12, 2015 to January 19, 2016, which contain a total coincident analysis time of 51.5 days accumulated when both detectors were operating in their normal state. As discussed in Ref. [13] with regard to the first 16 days of O1 data, the output data of both detectors typically contain nonstationary and non-Gaussian features, in the form of transient noise artifacts of varying durations. Longer duration artifacts, such as nonstationary behavior in the interferometer noise, are not very detrimental to CBC searches as they occur on a time scale that is much longer than any CBC waveform. However, shorter duration artifacts can pollute the noise background distribution of CBC searches. Many of these artifacts have distinct signatures [49] visible in the auxiliary data channels from the large number of sensors used to monitor instrumental or environmental disturbances at each observatory site [50]. When a significant noise source is identified, contaminated data are removed from the analysis data set. After applying this data quality process, detailed in Ref. [51], the remaining coincident analysis time in O1 is 48.6 days. The analyses search only stretches of data longer than a minimum duration, to ensure that the detectors are operating stably. The choice is different in the two analyses and reduces the available data to 46.1 days for the PyCBC analysis and 48.3 days for the GstLAL analysis.

## III. SEARCH RESULTS

Two different, largely independent, analyses have been implemented to search for stellar-mass BBH signals in the data of O1: PyCBC [2–4] and GstLAL [5–7]. Both these analyses employ matched filtering [52–60] with waveforms given by models based on general relativity [8,9] to search for gravitational waves from binary neutron stars, BBHs, and neutron star–black hole binaries. In this paper, we focus on the results of the matched-filter search for BBHs.

Results of the searches for binary neutron stars and neutron star–black hole binaries are reported in Ref. [43]. These matched-filter searches are complemented by generic transient searches which are sensitive to BBH mergers with total mass of about $30M_\odot$ or greater [61].

A bank of template waveforms is used to cover the parameter space to be searched [54,62–65]. The gravitational waveforms depend upon the masses $m_{1,2}$ (using the convention that $m_1 \geq m_2$) and angular momenta $S_{1,2}$ of the binary components. We characterize the angular momentum in terms of the dimensionless spin magnitude

$$a_{1,2} = \frac{c}{Gm_{1,2}^2}|S_{1,2}|, \quad (2)$$

and the component aligned with the direction of the orbital angular momentum, $L$, of the binary [66,67],

$$\chi_{1,2} = \frac{c}{Gm_{1,2}^2} S_{1,2} \cdot \hat{L}. \quad (3)$$

We restrict this template bank to circular binaries for which the spin of the systems is aligned (or antialigned) with the orbital angular momentum of the binary. The resulting templates can nonetheless recover systems with misaligned spins, which will exhibit orbital precession, with good sensitivity over much of the parameter space, particularly for near equal-mass binaries [44].

At leading order, the phase evolution during inspiral depends on the chirp mass of the system [68–70]

$$\mathcal{M} = \frac{(m_1 m_2)^{3/5}}{M^{1/5}}. \quad (4)$$

At subsequent orders in the PN expansion, the phase evolution depends predominantly upon the mass ratio [19]

$$q = \frac{m_2}{m_1} \leq 1, \quad (5)$$

and the effective spin parameter [71–76]

$$\chi_{\rm eff} = \frac{m_1\chi_1 + m_2\chi_2}{M}, \quad (6)$$

where $M = m_1 + m_2$ is the binary's total mass. The minimum black hole mass is taken to be $2M_\odot$, consistent with the largest known masses of neutron stars [77]. There is no known maximum black hole mass [78]; however, we limit this template bank to binaries with a total mass less than $M \leq 100M_\odot$. For higher-mass binaries, the Advanced LIGO detectors are sensitive to only the final few cycles of inspiral plus merger, making the analysis more susceptible to noise transients. The results of searches for more massive BBH mergers will be reported in future publications. In principle, black hole spins can lie anywhere in the range from $-1$ (maximal and antialigned) to $+1$ (maximal and aligned). We limit the spin magnitude to less than 0.9895, which is the region over which the EOBNR waveform model [8,9] used in the search is able to generate valid template waveforms [8]. The bank of templates used for the analysis is shown in Fig. 2.

Both analyses separately correlate the data from each detector with template waveforms that model the expected signal. The analyses identify candidate events that are detected at both the Hanford and Livingston observatories consistent with the 10-ms intersite propagation time. Additional signal consistency tests are performed to mitigate the effects of nonstationary transients in the data. Events are assigned a detection-statistic value that ranks their likelihood of being a gravitational-wave signal. For PyCBC, the observed SNR in each detector is reweighted using the signal consistency tests. These reweighted SNRs are added in quadrature to obtain the detection statistic $\hat{\rho}_c$. For GstLAL, $\ln \mathcal{L}$ is the log-likelihood ratio for the signal and noise models. The detection statistics are compared to the estimated detector noise background to determine, for each candidate event, the probability that detector noise would give rise to at least one equally significant event. Further details of the analysis methods are available in Appendix A.

The results for the two different analyses are presented in Fig. 3. The figure shows the observed distribution of events, as well as the background distribution used to assess significance. In both analyses, there are three events that lie above the estimated background: GW150914, GW151226, and LVT151012. All three of these are consistent with being BBH merger signals and are discussed in further detail below. The templates producing the highest significance in the two analyses are indicated in Fig. 2, the gravitational waveforms are shown in Fig. 1, and key parameters are summarized in Table I. There were no other significant BBH candidates in the first advanced LIGO observing run. All other observed events are consistent with the noise background for the search. A follow-up of the coincident events $\hat{\rho}_c \approx 9$ in the PyCBC analysis suggests that they are likely due to noise fluctuations or poor data quality, rather than a population of weaker gravitational-wave signals.

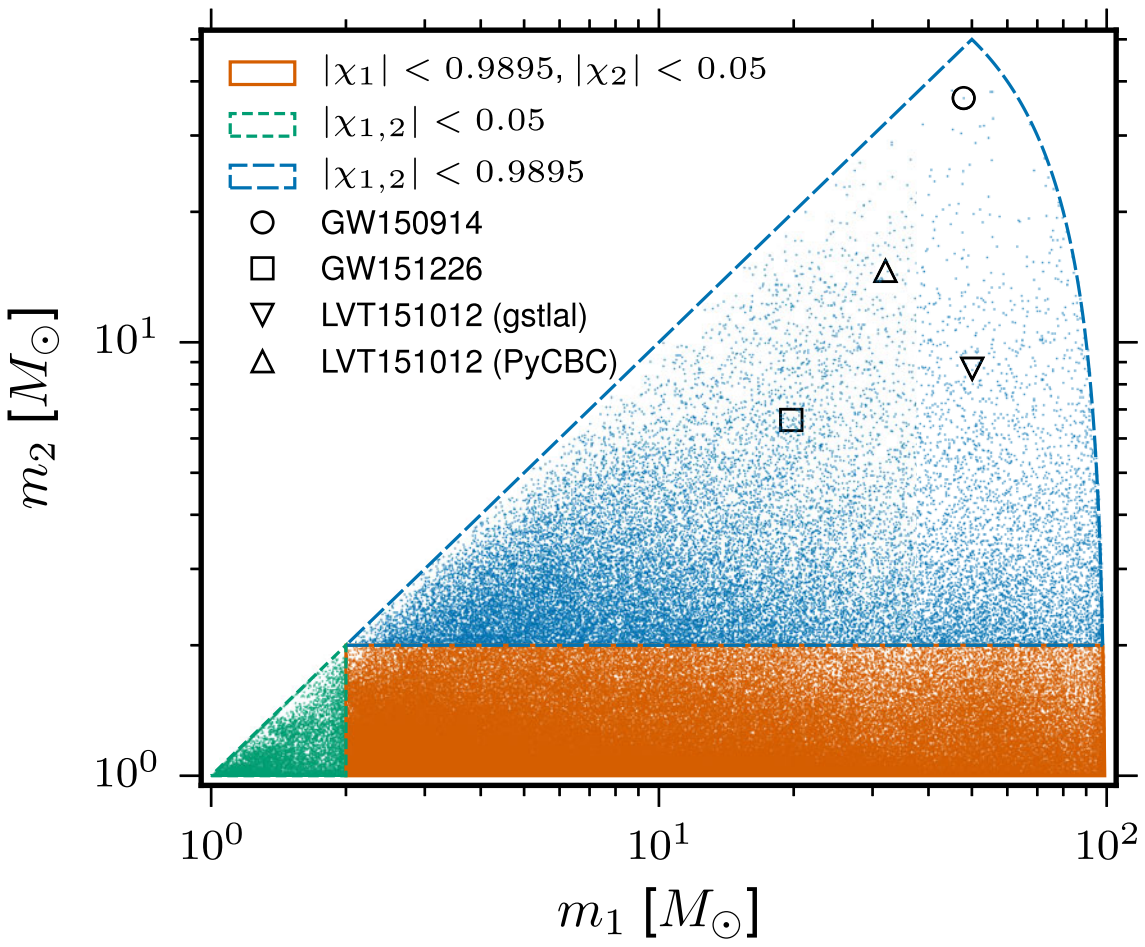

FIG. 2. The four-dimensional search parameter space covered by the template bank shown projected into the component-mass plane, using the convention $m_1 > m_2$. The colors indicate mass regions with different limits on the dimensionless spin parameters $\chi_1$ and $\chi_2$. Symbols indicate the best matching templates for GW150914, GW151226, and LVT151012. For GW150914 and GW151226, the templates were the same in the PyCBC and GstLAL searches, while for LVT151012 they differed. The parameters of the best matching templates are consistent, up to the discreteness of the template bank, with the detector frame mass ranges provided by detailed parameter estimation in Sec. IV.

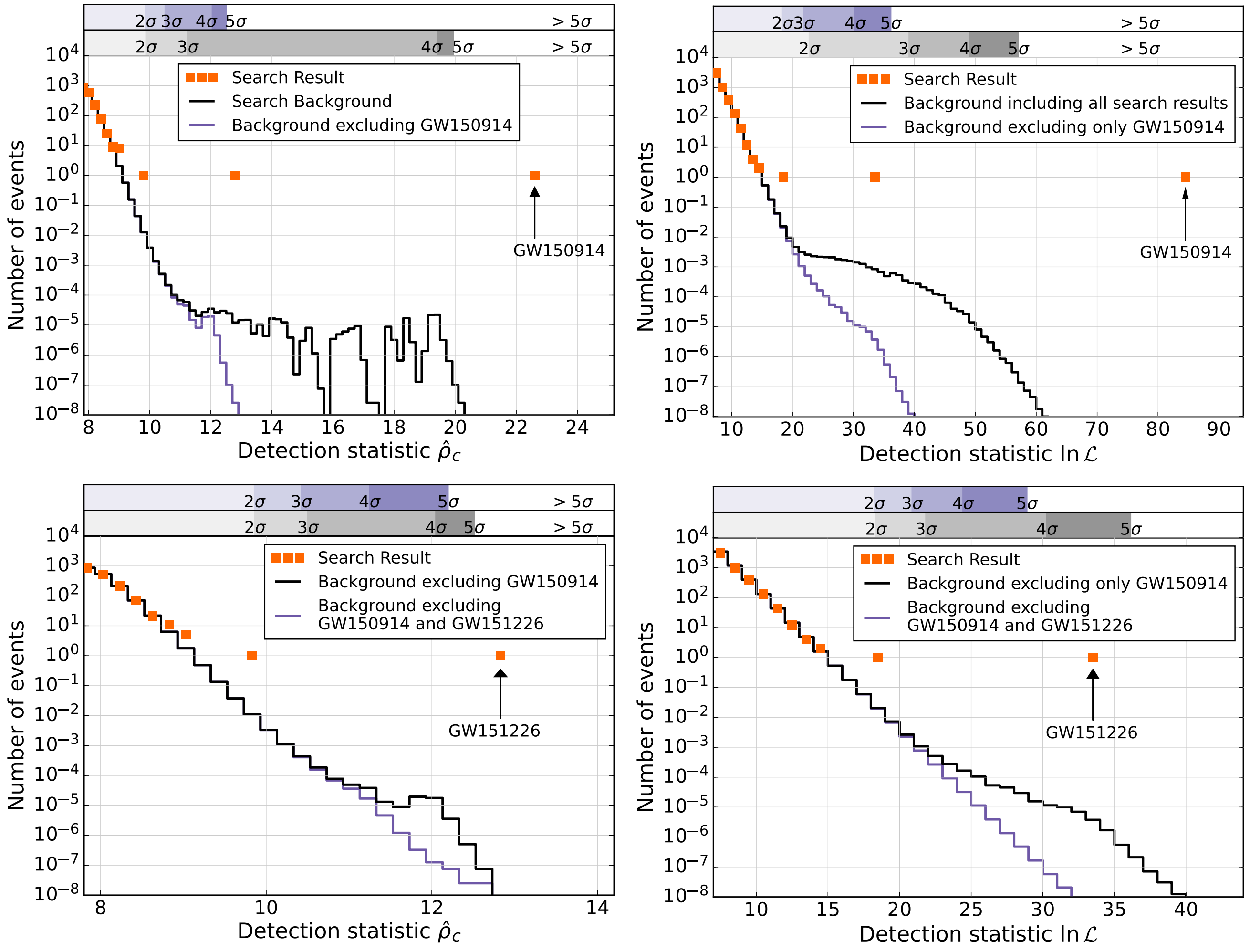

FIG. 3. Search results from the two analyses. The upper left-hand plot shows the PyCBC result for signals with chirp mass $\mathcal{M} > 1.74\ M_\odot$ (the chirp mass of an $m_1 = m_2 = 2\ M_\odot$ binary) and $f_{\rm peak} > 100$ Hz, while the upper right-hand plot shows the GstLAL result. In both analyses, GW150914 is the most significant event in the data, and it is more significant than any background event in the data. It is identified with a significance greater than $5\sigma$ in both analyses. As GW150914 is so significant, the high significance background is dominated by its presence in the data. Once it has been identified as a signal, we remove it from the background estimation to evaluate the significance of the remaining events. The lower plots show results with GW150914 removed from both the foreground and background, with the PyCBC result on the left and the GstLAL result on the right. In both analyses, GW151226 is identified as the most significant event remaining in the data. GW151226 is more significant than the remaining background in the PyCBC analysis, with a significance of greater than $5\sigma$. In the GstLAL search, GW151226 is measured to have a significance of $4.5\sigma$. The third most significant event in the search, LVT151012, is identified with a significance of $1.7\sigma$ and $2.0\sigma$ in the two analyses, respectively. The significance obtained for LVT151012 is not greatly affected by including or removing background contributions from GW150914 and GW151226.

It is clear from Fig. 3 that at high significance, the background distribution is dominated by the presence of GW150914 in the data. Consequently, once an event has been confidently identified as a signal, we remove triggers associated with it from the background in order to get an accurate estimate of the noise background for lower amplitude events. The lower panel of Fig. 3 shows the search results with GW150914 removed from both the foreground and background distributions.

### A. GW150914

GW150914 was observed on September 14, 2015 at 09:50:45 UTC with a matched-filter SNR of 23.7 [79]. It is recovered with a reweighted SNR in the PyCBC analysis of $\hat{\rho}_c = 22.7$ and a log likelihood of 84.7 in the GstLAL analysis. A detailed discussion of GW150914 is given in Refs. [16,39,44], where it was presented as the most significant event in the first 16 days of Advanced LIGO observing. The results presented here differ from the previous ones in two ways: They make use of the full O1 data set, and they use the final instrumental calibration. Thus, while GW150914 remains the most significant in this search, the recovered SNR and significance of the event differ slightly from the previously reported values. In particular, for the PyCBC analysis, the event is recovered with slightly lower SNR than with the preliminary calibration and with a higher value of the $\chi^2$ signal consistency test in the H1 detector. This leads to a reduction of the detection statistic $\hat{\rho}_c$, from 23.6 in Ref. [16] to the current value of 22.7. Additionally, for the PyCBC analysis, a redefinition of the mass bins used to group templates with similar background caused the significance of GW150914 to be evaluated against a different background; for details see Appendix A 1. For the GstLAL analysis of the full O1 data set, a decrease in the background probability for GW150914 increased the log likelihood to 84.7 from the original value of 78.

GW150914 remains the most significant event in both analyses. Furthermore, in both cases, there are no background events with significance equal to or greater than GW150914. Consequently, we can only calculate a limit on the false alarm rate (FAR) for GW150914. Using the time-shift method to estimate background, we limit the FAR of GW150914 to be less than $6.0 \times 10^{-7}$ yr$^{-1}$. This corresponds to a p-value of $7.5 \times 10^{-8}$, or a significance of $5.3\sigma$. The significance is greater than the $5.1\sigma$ derived in Ref. [44] due to a tripling of the analysis time, which allows time shifts to probe smaller false alarm rates.

The GstLAL analysis estimates the p-value assuming that noise triggers are equally likely to occur in any of the templates within a background bin. Under this assumption, the p-value of GW150914 is estimated to be $8.8 \times 10^{-12}$, which is the minimum p-value that can be informed by the data. However, as stated in Ref. [44], breaking that assumption implies that the minimum p-value would be higher. For this reason, we quote the more conservative PyCBC bound on the false alarm rate of GW150914 here and in Ref. [16].

### B. GW151226

GW151226 was observed on December 26, 2015 at 03:38:53 UTC with a combined matched-filter SNR of 13.0. The signal was identified as the second most significant event in both the PyCBC and GstLAL analyses with $\hat{\rho}_c = 12.8$ and $\ln \mathcal{L} = 22.6$, respectively.

Signal consistency tests show no sign of transient noise affecting the analyses at this time, and checks of the instrumental data reveal no serious data quality issues at the time of the event. When single interferometer triggers from GW150914 are used in our background estimation methods, the tail of the distribution is dominated by their presence. As GW150914 is confidently identified as a gravitational-wave signal [16], we remove any background events associated with it from the distribution.

The background distribution, under the assumption that GW150914 is a gravitational wave, is shown in the bottom row of Fig. 3. Now, GW151226 is more significant than all background events in the PyCBC analysis. Its significance cannot be measured and, as for GW150914, we limit the FAR to be less than $6.0 \times 10^{-7}$ yr$^{-1}$. This corresponds to a p-value of $7.5 \times 10^{-8}$, or a significance of $5.3\sigma$. In the GstLAL analysis, the background extends past the observed log likelihood of GW151226, and the event is recovered with a FAR of 1 per 44000 years, which corresponds to a p-value of $3.5 \times 10^{-6}$ and a significance of $4.5\sigma$.

### C. LVT151012

The third most significant event in O1 is LVT151012 observed on October 12, 2015 at 09:54:43 UTC. It was observed with a combined matched-filter SNR of 9.7 and detection statistic values $\hat{\rho}_c = 9.7$ and $\ln \mathcal{L} = 18.1$. The SNR of this event is considerably lower than GW150914 and GW151226 and, even though the signal consistency tests show no signs of noise origin, the search background is such that the FAR of LVT151012 is 1 per 2.7 years and 1 per 5.9 years in the PyCBC and GstLAL analyses, respectively. This equates to p-values of 0.045 and 0.025, or significances of $1.7\sigma$ and $2.0\sigma$. These results are consistent with expectations for candidate events with low matched-filter SNR since PyCBC and GstLAL use different ranking statistics and background estimation methods. At the significance of LVT151012, the background has contributions from a large number of triggers in each detector and is no longer dominated by the presence of GW150914 and GW151226 in the data. Consequently, removing them does not have a large effect on the significance. For PyCBC, the estimate of the significance is essentially unaffected by the removal of the events. For GstLAL, inclusion of GW150914 changes the p-value of LVT151012 by a factor of 2, but inclusion of GW151226 has little effect.

The significance of LVT151012 is such that we do not confidently claim this event as a gravitational-wave signal. However, it is more likely to be a gravitational-wave signal than noise based on our estimate for the rate of gravitational-wave signals (see Sec. VI). Detector characterization studies have not identified an instrumental or environmental artifact as causing this candidate event [13]. Parameter-estimation results for LVT151012 are presented in the following section and are consistent with our expectations for an astrophysical BBH source. The inferred component masses of LVT151012 lie roughly between the masses of GW150914 and GW151226, as shown in Fig. 4.

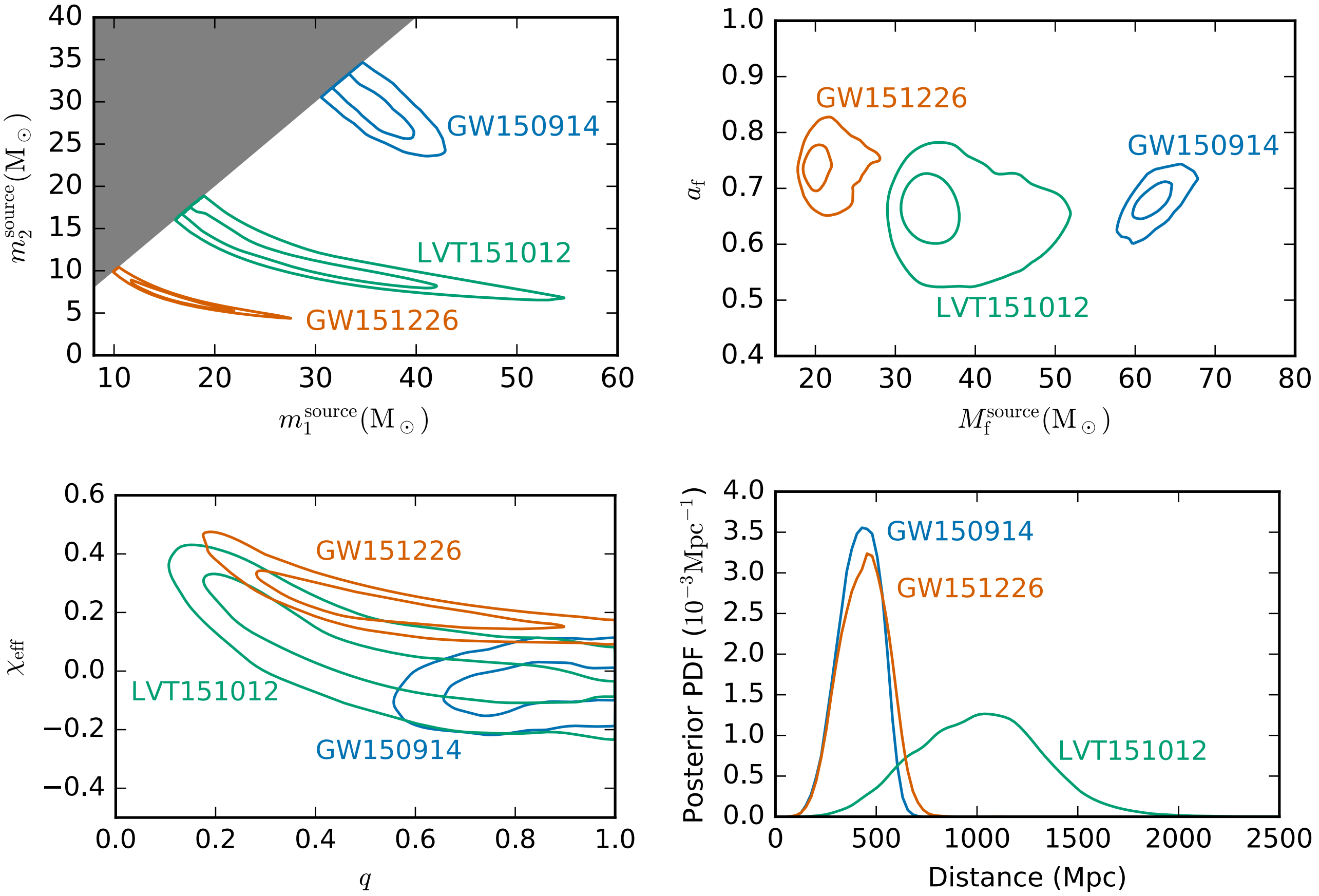

FIG. 4. Posterior probability densities of the masses, spins, and distance to the three events GW150914, LVT151012, and GW151226. For the two-dimensional distributions, the contours show 50% and 90% credible regions. Top left panel: Component masses $m_1^{\rm source}$ and $m_2^{\rm source}$ for the three events. We use the convention that $m_1^{\rm source} \geq m_2^{\rm source}$, which produces the sharp cut in the two-dimensional distribution. For GW151226 and LVT151012, the contours follow lines of constant chirp mass ($\mathcal{M}^{\rm source} = 8.9^{+0.3}_{-0.3}\mathrm{M}_\odot$ and $\mathcal{M}^{\rm source} = 15.1^{+1.4}_{-1.1}\mathrm{M}_\odot$, respectively). In all three cases, both masses are consistent with being black holes. Top right panel: The mass and dimensionless spin magnitude of the final black holes. Bottom left panel: The effective spin and mass ratios of the binary components. Bottom right panel: The luminosity distance to the three events.

## IV. SOURCE PROPERTIES

In this section, we present the inferred properties of the sources of GW150914, LVT151012, and GW151226, assuming that the signals each originate from a binary coalescence as described by general relativity. Tests of the consistency of the signal with the predictions of general relativity are presented in Sec. V. Full results for GW150914 have been provided in Refs. [39,40], and key results for LVT151012 have been given in Ref. [44]. Here, we give results based upon an updated calibration of the data. The analyses of all three signals closely mirror the original analysis of GW150914, as detailed in Ref. [39] and described in Appendix B.

The analysis makes use of two waveform models, the double aligned spin waveform model (EOBNR) [8,9] and an effective precessing spin model (IMRPhenom) [36–38]. Results from the two waveforms are similar, and the data give us little reason to prefer one model over the other. We therefore average the posterior distributions from two waveforms for our overall results. These are used for the discussion below, except in Sec. IV B, where we also consider measurements of spin alignment from the precessing IMRPhenom waveform.

The results match our expectations for a coherent signal in both detectors and give us no reason to suspect that any of the signals are not of astrophysical origin. All three signals are consistent with originating from BBHs. Key parameters for the three events are included in Table I and plotted in Figs. 4,5, and 6. Detailed results are provided in Table IV in Appendix B.

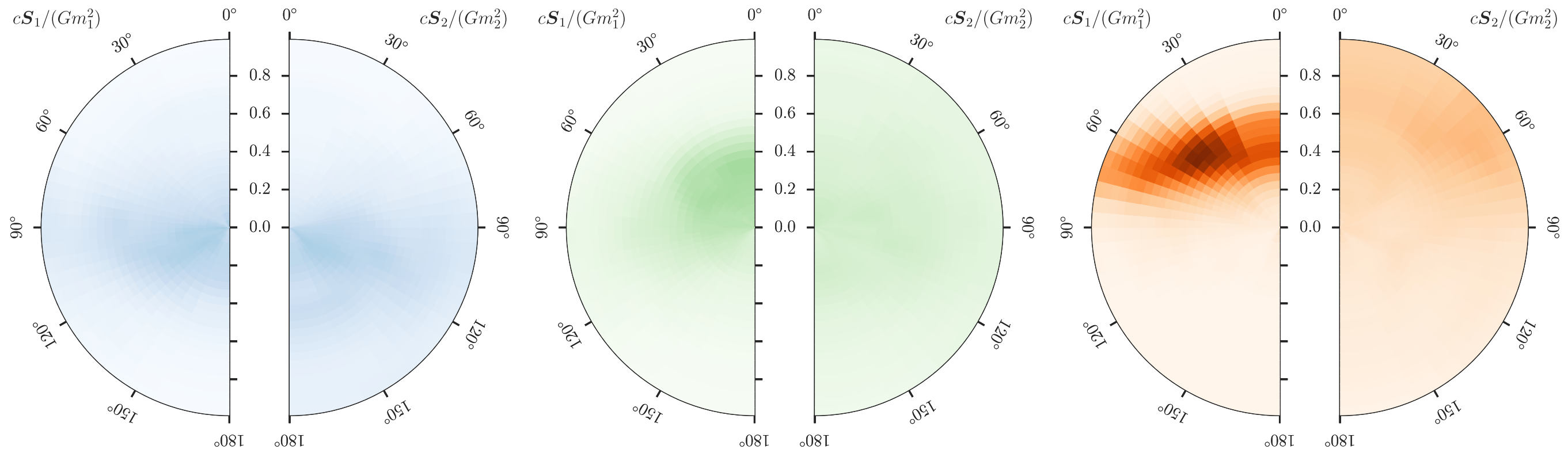

FIG. 5. Posterior probability distributions for the dimensionless component spins $cS_1/(Gm_1^2)$ and $cS_2/(Gm_2^2)$ relative to the normal to the orbital plane $L$, marginalized over the azimuthal angles. The bins are constructed linearly in spin magnitude and the cosine of the tilt angles, and therefore have equal prior probability. The left plot shows the distribution for GW150914, the middle plot is for LVT151012, and the right plot is for GW151226.

### A. Masses

The binary component masses of all three systems lie within the range expected for stellar-mass black holes. The least massive black hole is the secondary of GW151226, which has a 90% credible lower bound that $m_2^{\rm source} \geq 5.6 M_\odot$. This is above the expected maximum neutron star mass of about $3 M_\odot$ [80,81] and beyond the mass gap where there is currently a dearth of black holes observed in x-ray binaries [82–84]. The range of our inferred component masses overlaps with those for stellar-mass black holes measured through x-ray observations but extends beyond the nearly $16 M_\odot$ maximum of that population [85–87].

GW150914 corresponds to the heaviest BBH system ($M^{\rm source} = 65.3^{+4.1}_{-3.4} M_\odot$) we observed, and GW151226 corresponds to the least massive ($M^{\rm source} = 21.8^{+5.9}_{-1.7} M_\odot$). Higher mass systems merge at a lower gravitational-wave frequency. For lower-mass systems, the gravitational-wave signal is dominated by the inspiral of the binary components, whereas for higher-mass systems, the merger and ringdown parts of the signal are increasingly important. The transition from being inspiral dominated to being merger and ringdown dominated depends upon the sensitivity of the detector network as a function of frequency; GW150914 had SNR approximately equally split between the inspiral and post-inspiral phases [41]. Information about the masses is encoded in different ways in the different parts of the waveform: The inspiral predominantly constrains the chirp mass [70,88,89], and the ringdown is more sensitive to the total mass [90]; hence, the best-measured parameters depend upon the mass [91–93]. This is illustrated in the posterior probability distributions for the three events in Fig. 4. For the lower-mass GW151226 and LVT151012, the posterior distribution follows curves of constant chirp mass, but for GW150914, the posterior is shaped more by constraints on the total mass [94].

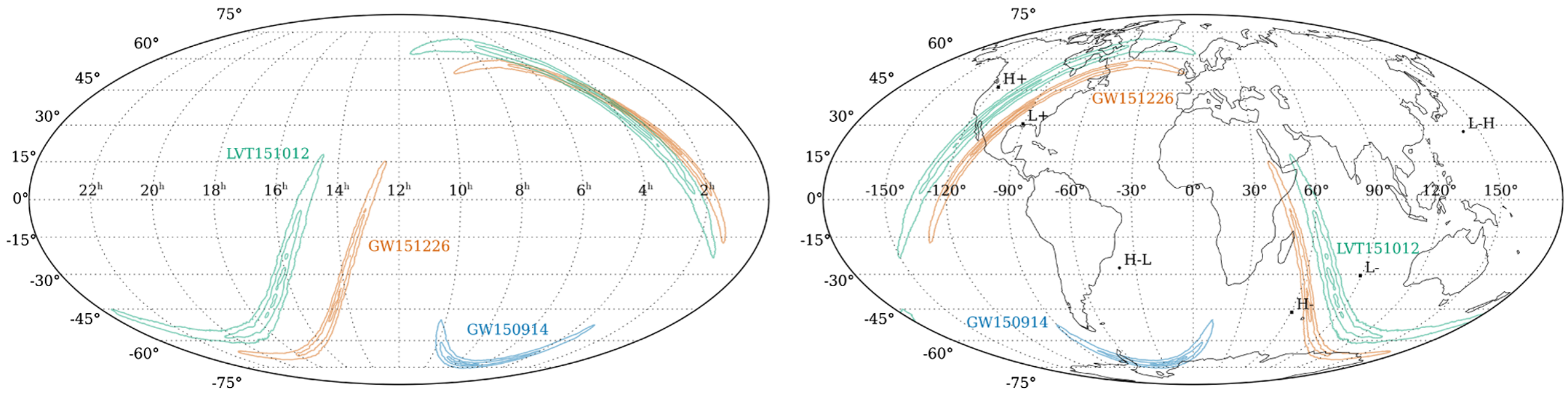

FIG. 6. Posterior probability distributions for the sky locations of GW150914, LVT151012, and GW151226 shown in a Mollweide projection. The left plot shows the probable position of the source in equatorial coordinates (right ascension is measured in hours and declination is measured in degrees). The right plot shows the localization with respect to the Earth at the time of detection. H+ and L+ mark the Hanford and Livingston sites, and H− and L− indicate antipodal points; H-L and L-H mark the poles of the line connecting the two detectors (the points of maximal time delay). The sky localization forms part of an annulus, set by the difference in arrival times between the detectors.

The mass ratio $q$ also differs between the events. We infer that GW150914 came from a near equal-mass system (the 90% credible lower bound of the mass ratio is $q \geq 0.65$), but GW151226 and LVT151012 have posterior support for more unequal-mass ratios ($q \geq 0.28$ and $q \geq 0.24$, respectively). The mass ratio has a large uncertainty, as it is degenerate with the spin of the compact objects [89,95,96]. This degeneracy could be broken if a signal contains a clear imprint of precession [97–100], but we have yet to observe this signature (see Sec. IV B). Measurement of the mass ratio could inform our understanding of the origin of BBH systems.

Following the inspiral, the BBHs merge to form a final remnant black hole. We estimate the masses of these using fitting formulas calibrated to numerical relativity simulations [36,101]. Each final mass is 0.95–0.98 of the initial total mass of the binary components, as similar fractions of 0.02–0.05 are radiated away as gravitational waves. While predominantly determined by the total mass, the radiated energy also depends upon the mass ratio and component spins; our results are consistent with expectations for moderately spinning black holes [102,103]. The remnant black holes are more massive than any black hole observed to date in an x-ray binary, the least massive being GW151226's $M_\mathrm{f}^\mathrm{source} = 20.8^{+6.1}_{-1.7}\mathrm{M}_\odot$. The final black hole masses, as well as their spins, are shown in Fig. 4. The remnant for GW150914 has a mass of $M_\mathrm{f}^\mathrm{source} = 62.3^{+3.7}_{-3.1}\mathrm{M}_\odot$ and is the most massive stellar-mass black hole observed to date.

BBH mergers have extremely high gravitational-wave luminosities: The peak values are $3.6^{+0.5}_{-0.4} \times 10^{56}\ \mathrm{erg\,s^{-1}}$, $3.1^{+0.8}_{-1.8} \times 10^{56}\ \mathrm{erg\,s^{-1}}$, and $3.3^{+0.8}_{-1.6} \times 10^{56}\ \mathrm{erg\,s^{-1}}$ for GW150914, LVT151012, and GW151226, respectively. These luminosities are calculated using a fit to nonprecessing numerical-relativity simulations [104], and the uncertainty includes the estimated error from this fit. Whereas the energy radiated scales with the total mass, the luminosity is comparable for all three systems. There is some variation from differences in the mass ratios and spins, and uncertainties in these dominate the overall uncertainty. The luminosity is independent of the total mass, as this sets both the characteristic energy scale and characteristic time scale for the system [105].

### B. Spins

An isolated black hole has three intrinsic properties: mass, spin, and electric charge [106–109]. We expect the charge of astrophysical black holes to be negligible [110–112]. Both the masses and spins of the black holes leave an imprint on the gravitational-wave signal during a coalescence. The components of the spins parallel to the orbital angular momentum affect the phasing of the binary, whereas orthogonal components lead to orbital precession. The effects of the spins of the binary components are subdominant, and they are more difficult to constrain than the masses.

Only weak constraints can be placed on the spin magnitudes of the binary components: In all cases, the uncertainty spans the majority of the allowed range of [0, 1]. We can better infer the spin of the more massive black hole, as this has a greater impact upon the inspiral. We find that smaller spins are favored, and we place 90% credible bounds on the primary spin $a_1 \leq 0.7$ for GW150914 and LVT151012, and $a_1 \leq 0.8$ for GW151226.

Observations for all three events are consistent with small values of the effective spin: $|\chi_\mathrm{eff}| \leq 0.17$, 0.28, and 0.35 at 90% probability for GW150914, LVT151012, and GW151226, respectively. This result indicates that large parallel spins aligned or antialigned with the orbital angular momentum are disfavored. Only in the case of GW151226 do we infer a nonzero value of $\chi_\mathrm{eff}$, and from this, we infer that at least one of the components has a spin of $\geq 0.2$ at the 99% credible level.

Misalignment of the component spins with respect to the orbital angular momentum leads to precession [113]. As a first approximation, the amount of precession may be quantified through a single effective precession spin parameter $\chi_\mathrm{p}$ [114]. The inferred distributions for $\chi_\mathrm{p}$ are roughly consistent with our prior expectations after incorporating the measured constraints on $\chi_\mathrm{eff}$. The absence of clear information about precession could be because there is intrinsically little precession since the binary is orientated nearly face-on or face-off (see Sec. IV C), which minimizes the visible effect of precession, or because of a combination of these effects. Our aligned-spin search has reduced sensitivity to highly precessing systems [44], which makes it more probable that we detect nonprecessing systems. We have yet to find strong evidence for precession but cannot exclude the possibility of misaligned spins.

The posterior probabilities for the spin magnitudes and tilts relative to the orbital angular momentum using the precessing IMRPhenom model are shown in Fig 5. In all cases, larger spin magnitudes are allowed when the spin is misaligned: The additional in-plane spin does not change $\chi_\mathrm{eff}$. For LVT151012 and GW151226, there is significantly greater uncertainty for the spin of the secondary than for the primary. This is because the mass ratios for these systems can be more extreme: For equal mass binaries, both spins play an equal role in the dynamics, but, as the mass ratio tends towards zero, the effects of the secondary spin become negligible.

All three events have final black holes with spins of about 0.7, as expected for mergers of similar-mass black holes [115,116]. The final spin is dominated by the orbital angular momentum of the binary at merger. Consequently, it is more precisely constrained than the component spins and is broadly similar across the three events. The masses and spins of the final black holes are plotted in Fig. 4.

The spin of the final black hole, like its mass, is calculated using fitting formulas calibrated against numerical relativity simulations. In Ref. [39], we used a formula that only included contributions from the aligned components of spins [101]; we now use an extension of this formula, which also incorporates the effects of in-plane spins [117]. The change has a small impact on the final spin of GW150914 (changing from $0.67^{+0.05}_{-0.06}$ to $0.68^{+0.05}_{-0.06}$) and a larger effect on GW151226 (changing from $0.72^{+0.05}_{-0.05}$ to $0.74^{+0.06}_{-0.06}$) as its components have more significant spins.

### C. Distance, inclination, and sky location

The luminosity distance to the source is inversely proportional to the signal's amplitude. GW150914 and GW151226 have comparable distance estimates of $D_\mathrm{L} = 420^{+150}_{-180}$ Mpc (redshift $z = 0.09^{+0.03}_{-0.04}$) and $D_\mathrm{L} = 440^{+180}_{-190}$ Mpc ($z = 0.09^{+0.03}_{-0.04}$), respectively [118]. GW151226 originates from a lower-mass system than GW150914; hence, the gravitational-wave signal is intrinsically quieter, and its SNR is lower than GW150914's even though the distances are comparable. LVT151012 is the quietest signal and is inferred to be at a greater distance $D_\mathrm{L} = 1000^{+500}_{-500}$ Mpc ($z = 0.20^{+0.09}_{-0.09}$).

In all cases, there is significant fractional uncertainty for the distance. This is predominantly a consequence of the degeneracy between the distance and the binary's inclination, which also impacts the signal amplitude [95,119,120].

The inclination is only weakly constrained; in all cases, there is greatest posterior support for the source being either face-on or face-off (angular momentum pointed parallel or antiparallel to the line of sight). This is the orientation that produces the greatest gravitational-wave amplitude, so it is consistent with the largest distance. The inclination could potentially be better constrained in a precessing system [98,121]. Only for GW150914 is there preference for one of the configurations, with there being greater posterior support for the source being face-off [39].

Sky localization from a gravitational-wave detector network is primarily determined by the measured delay in the signal arriving at the sites, with additional information coming from the signal amplitude and phase [122–124]. For a two-detector network, the sky localization forms a characteristic broken annulus [125–128]. Adding additional detectors to the network would improve localization abilities [129–132]. The sky localizations of the three events are shown in Fig. 6, including both celestial coordinates (indicating the origin of the signal) and geographic coordinates (illustrating localization with respect to the two detectors). The arrival time at Hanford relative to Livingston was $\Delta t_\mathrm{HL} = 7.0^{+0.2}_{-0.2}$ ms for GW150914, $\Delta t_\mathrm{HL} = -0.6^{+0.6}_{-0.6}$ ms for LVT151012, and $\Delta t_\mathrm{HL} = 1.1^{+0.3}_{-0.3}$ ms for GW151226. Both LVT151012 and GW151226 are nearly overhead of the two detectors, which is where we are most sensitive and hence expect to make most detections [53,133].

The 90% credible region for sky localization is 230 $\mathrm{deg}^2$ for GW150914, 850 $\mathrm{deg}^2$ for GW151226, and 1600 $\mathrm{deg}^2$ for LVT151012. As expected, the sky area is larger for quieter events. The sky area is expected to scale inversely with the square of the SNR [128,134], and we see that this trend is followed.

## V. TESTS OF GENERAL RELATIVITY

GW150914 provided us with the first empirical access to the genuinely strong field dynamics of gravity. With the frequency of the waveform peak amplitude well aligned with the best instrument sensitivity, the late inspiral and merger-ringdown regime could be studied in considerable detail, as described in Ref. [41]. This allows for checks of the consistency between masses and spins estimated from different portions of the waveform [135], as well as parametrized tests of the waveform as a whole [136–139]. Even though not much of the early inspiral was in the detectors' sensitive band, interesting bounds were placed on departures from general relativity in the PN coefficients up to 3.5PN. Since the source of GW151226 merged at about 450 Hz, the signal provides the opportunity to probe the PN inspiral with many more waveform cycles, albeit at relatively low SNR. Especially in this regime, GW151226 allows us to further tighten our bounds on violations of general relativity.

As in Ref. [41], to analyze GW151226, we start from the IMRPhenom waveform model of Refs. [36–38], which is capable of describing inspiral, merger, and ringdown, and partly accounts for spin precession. The phase of this waveform is characterized by coefficients $\{p_i\}$, which include PN coefficients, as well as phenomenological coefficients describing merger and ringdown. The latter were obtained by calibrating against numerical waveforms and tend to multiply specific powers of $f$. They characterize the gravitational-wave amplitude and phase in different stages of the coalescence process. We allow for possible departures from general relativity, parametrized by a set of testing coefficients $\delta\hat{p}_i$, which take the form of *fractional* deviations in the $p_i$ [140,141]. Thus, we replace $p_i$ by $(1 + \delta\hat{p}_i)p_i$ and let one or more of the $\delta\hat{p}_i$ vary freely, in addition to the source parameters that also appear in pure general relativity waveforms, using the general relativistic expressions for $p_i$ in terms of masses and spins. Our testing coefficients are those in Table I of Ref. [41]. For convenience, we list them again: (i) $\{\delta\hat{\varphi}_0, \ldots, \delta\hat{\varphi}_7\}$ [142] and $\{\delta\hat{\varphi}_{5l}, \delta\hat{\varphi}_{6l}\}$ for the PN coefficients (where the last two multiply a term of the form $f^\gamma \log f$), (ii) intermediate-regime parameters $\{\delta\hat{\beta}_2, \delta\hat{\beta}_3\}$, and (iii) merger-ringdown parameters $\{\delta\hat{\alpha}_2, \delta\hat{\alpha}_3, \delta\hat{\alpha}_4\}$ [143].

In our analyses, we let each one of the $\delta\hat{p}_i$ in turn vary freely, while all others are fixed to their general relativity values, $\delta\hat{p}_j = 0$ for $j \neq i$. These tests model general relativity violations that would occur predominantly at a particular PN order (or in the case of the intermediate and merger-ringdown parameters, a specific power of frequency in the relevant regime), although together they can capture deviations that are measurably present at more than one order.

In Ref. [41], for completeness, we have also shown results from analyses where the parameters in each of the regimes (i)–(iii) are allowed to vary simultaneously, but these tests return wide and uninformative posteriors. By contrast, analyses where the testing parameters $\delta\hat{p}_i$ are varied one at a time have much smaller statistical uncertainties. Moreover, as demonstrated in Ref. [144], checking for a deviation from zero in a single testing parameter is an efficient way to uncover GR violations that occur at multiple PN orders, and one can even find violations at powers of frequency that are distinct from the one that the testing parameter is associated with [145,146]. Hence, such analyses are well suited to search for generic departures from GR, though it should be stressed that if a violation is present, the measured values of the $\delta\hat{p}_i$ will not necessarily reflect the predicted values of the correct alternative theory. To reliably constrain theory-specific quantities such as coupling constants or extra charges, one should directly apply full inspiral-merger-ringdown waveform models from specific modified gravity theories [147], but in most cases, these are not yet available. However, in the present work, the focus is on model-independent tests of general relativity itself.

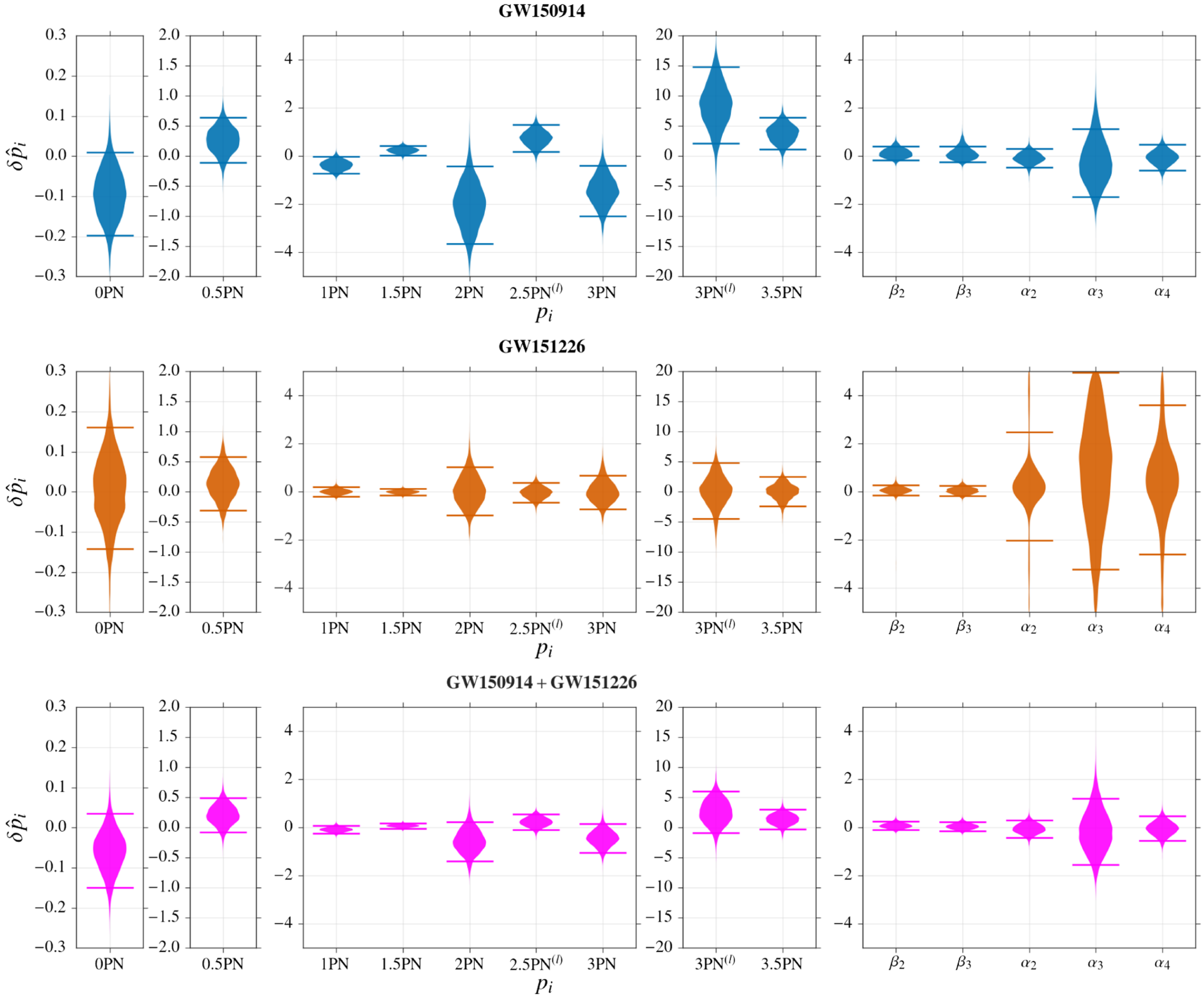

FIG. 7. Posterior density distributions and 90% credible intervals for relative deviations $\delta\hat{p}_i$ in the PN parameters $p_i$ (where $(l)$ denotes the logarithmic correction), as well as intermediate parameters $\beta_i$ and merger-ringdown parameters $\alpha_i$. The top panel is for GW150914 by itself and the middle one for GW151226 by itself, while the bottom panel shows *combined* posteriors from GW150914 and GW151226. While the posteriors for deviations in PN coefficients from GW150914 show large offsets, the ones from GW151226 are well centered on zero, as well as being tighter, causing the combined posteriors to similarly improve over those of GW150914 alone. For deviations in the $\beta_i$, the combined posteriors improve over those of either event individually. For the $\alpha_i$, the joint posteriors are mostly set by the posteriors from GW150914, whose merger-ringdown occurred at frequencies where the detectors are the most sensitive.

Given the observation of more than one BBH merger, posterior distributions for the $\delta\hat{p}_i$ can be combined to yield stronger constraints. In Fig. 7, we show the posteriors from GW150914, generated with final instrumental calibration, and GW151226 by themselves, as well as joint posteriors from the two events together. We do not present similar results for the candidate LVT151012 since it is not as confident a detection as the others; furthermore, its smaller detection SNR means that its contribution to the overall posteriors is insignificant.

For GW150914, the testing parameters for the PN coefficients, $\delta\hat{\varphi}_i$ and $\delta\hat{\varphi}_{il}$, showed moderately significant ($2\sigma$–$2.5\sigma$) deviations from their general relativity values of zero [41]. By contrast, the posteriors of GW151226 tend to be centered on the general relativity value. As a result, the offsets of the combined posteriors are smaller. Moreover, the joint posteriors are considerably tighter, with a $1\sigma$ spread as small as 0.07 for deviations in the 1.5PN parameter $\varphi_3$, which encapsulates the leading-order effects of the dynamical self-interaction of spacetime geometry (the "tail" effect) [148–151], as well as spin-orbit interaction [67,152,153].

In Fig. 8, we show the 90% credible upper bounds on the magnitude of the fractional deviations in PN coefficients, $|\delta\hat{\varphi}_i|$, which are affected by both the offsets and widths of the posterior density functions for the $\delta\hat{\varphi}_i$. We show bounds for GW150914 and GW151226 individually, as well as the joint upper bounds resulting from the combined posterior density functions of the two events. Not surprisingly, the quality of the joint bounds is mainly due to GW151226 because of the larger number of inspiral cycles in the detectors' sensitive frequency band. Note how at high PN order, the combined bounds are slightly looser than the ones from GW151226 alone; this is because of the large offsets in the posteriors from GW150914.

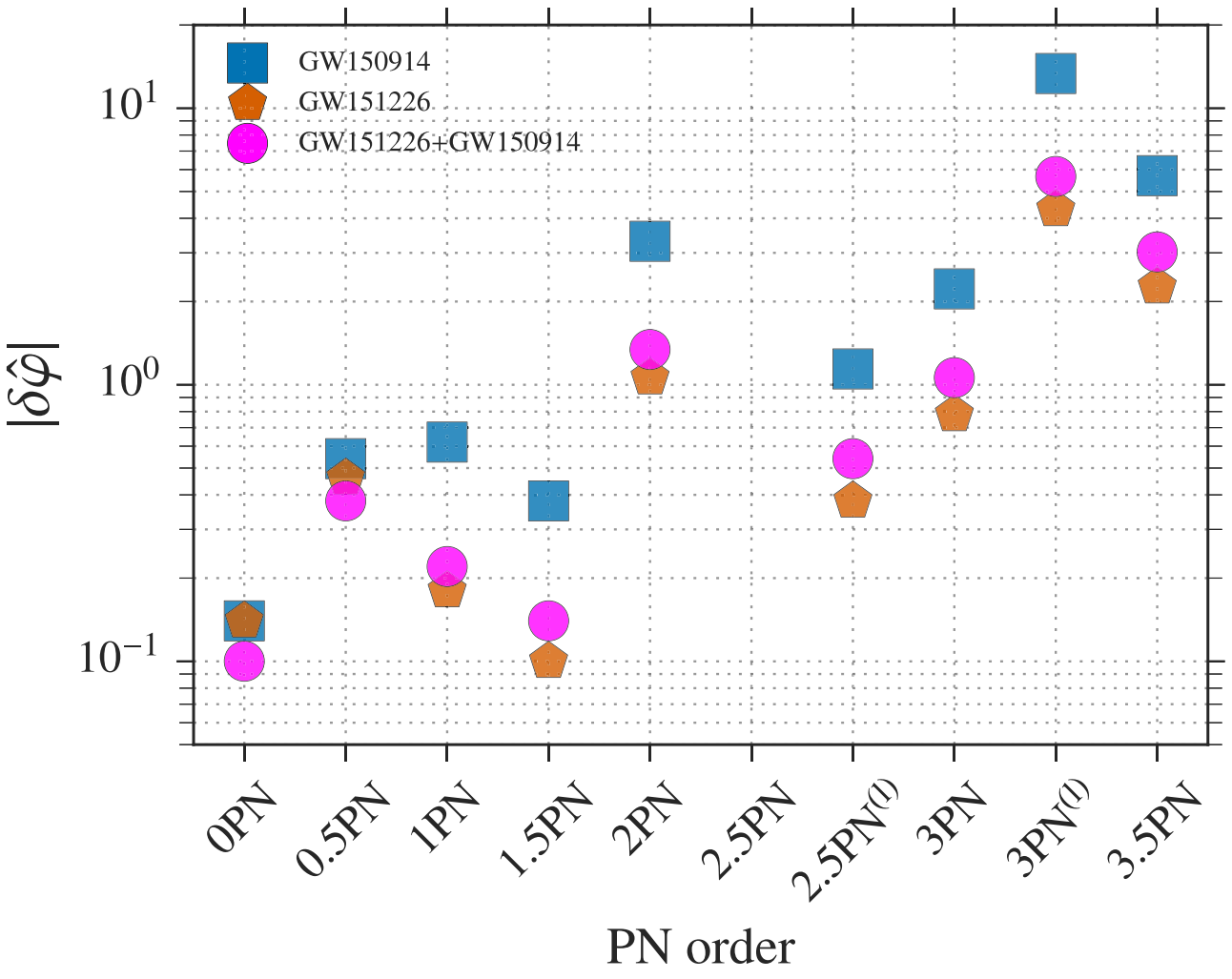

FIG. 8. The 90% credible upper bounds on deviations in the PN coefficients, from GW150914 and GW151226. Also shown are joint upper bounds from the two detections; the main contributor is GW151226, which had many more inspiral cycles in band than GW150914. At 1PN order and higher, the joint bounds are slightly looser than the ones from GW151226 alone; this is due to the large offsets in the posteriors for GW150914.

Next, we consider the intermediate-regime coefficients $\delta\hat{\beta}_i$, which pertain to the transition between inspiral and merger-ringdown. For GW151226, this stage is well inside the sensitive part of the detectors' frequency band. Returning to Fig. 7, we see that the measurements for GW151226 are of comparable quality to GW150914, and the combined posteriors improve on the ones from either detection by itself. Last, we look at the merger-ringdown parameters $\delta\hat{\alpha}_i$. For GW150914, this regime corresponded to frequencies of $f \in [130, 300]$ Hz, while for GW151226, it occurred at $f \gtrsim 400$ Hz. As expected, the posteriors from GW151226 are not very informative for these parameters, and the combined posteriors are essentially determined by those of GW150914.

In summary, GW151226 makes its most important contribution to the combined posteriors in the PN inspiral regime, where both offsets and statistical uncertainties have significantly decreased over the ones from GW150914, in some cases almost to the 10% level.

An inspiral-merger-ringdown consistency test as performed on GW150914 in Ref. [41] is not meaningful for GW151226 since very little of the signal is observed in the post-merger phase. Likewise, the SNR of GW151226 is too low to allow for an analysis of residuals after subtraction of the most probable waveform. In Ref. [41], GW150914 was used to place a lower bound on the graviton Compton wavelength of $10^{13}$ km GW151226 gives a somewhat weaker bound because of its lower SNR, so combining information from the two signals does not significantly improve on this; an updated bound must await further observations. Finally, BBH observations can be used to test the consistency of the signal with the two polarizations of gravitational waves predicted by general relativity [154]. However, as with GW150914, we are unable to test the polarization content of GW151226 with the two, nearly aligned aLIGO detectors. Future observations, with an expanded network, will allow us to look for evidence of additional polarization content arising from deviations from general relativity.

## VI. BINARY BLACK HOLE MERGER RATES

The observations reported here enable us to constrain the rate of BBH coalescences in the local Universe more precisely than was achieved in Ref. [42] because of the longer duration of data containing a larger number of detected signals.

To do so, we consider two classes of triggers: those whose origin is astrophysical and those whose origin is terrestrial. Terrestrial triggers are the result of either instrumental or environmental effects in the detector, and their distribution is calculated from the search background estimated by the analyses (as shown in Fig. 3). The distribution of astrophysical events is determined by performing large-scale simulations of signals drawn from astrophysical populations and added to the data set. We then use our observations to fit for the number of triggers of terrestrial and astrophysical origin, as discussed in detail in Appendix C. The details of the astrophysical population have a minimal impact on the fit, as in all cases we assume a population distributed uniformly in comoving volume. Figure 9 shows the inferred distributions of signal and noise triggers, as well as the combined distribution. The observations are in good agreement with the model. GW150914 stands somewhat above the inferred distribution, as it is an unusually significant event—only 6% of the astrophysical population of sources appearing in our search with a false rate of less than one per century will be more significant than GW150914.

It is clear from the figure that three triggers are more likely to be signal (i.e., astrophysical) than noise (terrestrial). We evaluate this probability and find that, for GW150914 and GW151226, the probability of astrophysical origin is unity to within one part in $10^6$. Meanwhile, for LVT151012, it is calculated to be 0.87 and 0.86, for the PyCBC and GstLAL analyses, respectively. For all of the remaining events, the probability of astrophysical origin is less than 15%.

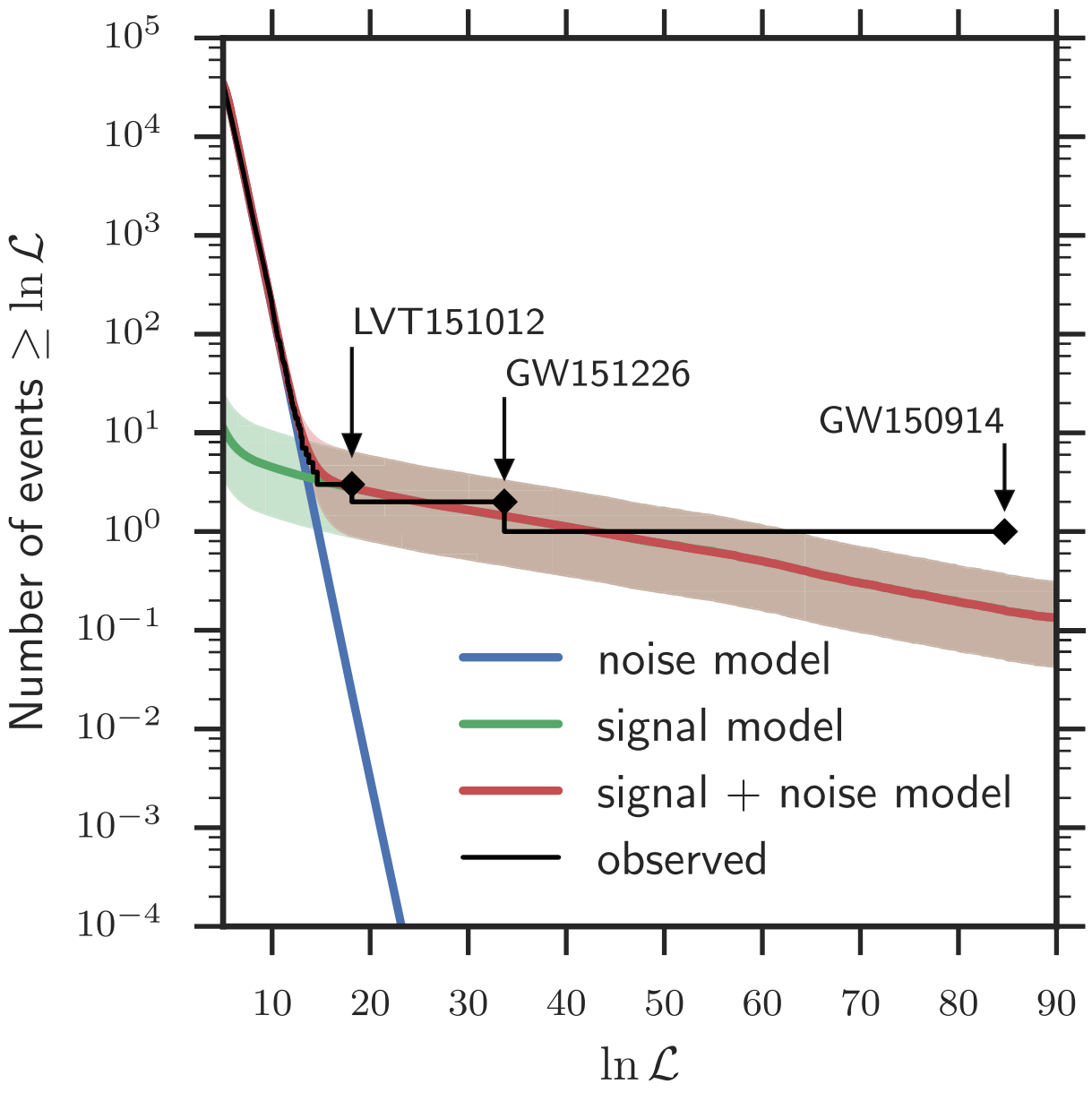

FIG. 9. The cumulative (right to left) distribution of observed triggers in the GstLAL analysis as a function of the log likelihood. The best-fit signal + noise distribution, and the contributions from signal and noise are also shown. The lines show the median number of expected triggers, and shaded regions show $1\sigma$ uncertainties. The observations are in good agreement with the model. At low likelihood, the distribution matches the noise model, while at high likelihood, it follows the signal model. Three triggers are clearly identified as being more likely to be signal than noise.

Given uncertainty in the formation channels of the various BBH events, we calculate the inferred rates using a variety of source population parametrizations. For a given population, the rate is calculated as $R = \Lambda/\langle VT\rangle$, where $\Lambda$ is the number of triggers of astrophysical origin and $\langle VT\rangle$ is the population-averaged sensitive space-time volume of the search. We use two canonical distributions for BBH masses:

(i) a distribution uniform (flat) over the logarithm of component masses, $p(m_1, m_2) \propto m_1{}^{-1} m_2{}^{-1}$ and
(ii) assuming a power-law distribution in the primary mass, $p(m_1) \propto m_1^{-2.35}$, with a uniform distribution on the second mass.

We require $5\mathrm{M}_\odot \leq m_2 \leq m_1$ and $m_1 + m_2 \leq 100\mathrm{M}_\odot$. The first distribution probably overestimates the fraction of high-mass black holes and therefore overestimates $\langle VT\rangle$, resulting in an underestimate of the true rate, while the second probably overestimates the fraction of low-mass black holes and therefore underestimating $\langle VT\rangle$ and overestimating the true rate. The inferred rates for these two populations are shown in Table II, and the rate distributions are plotted in Fig. 11.

In addition, we calculate rates based upon the inferred properties of the three significant events observed in the data: GW150914, GW151226, and LVT151012, as discussed in Appendix C. Since these classes are distinct, the total event rate is the sum of the individual rates: $R \equiv R_{\mathrm{GW150914}} + R_{\mathrm{LVT151012}} + R_{\mathrm{GW151226}}$. Note that the total rate estimate is dominated by GW151226, as it is the least massive of the three likely signals and is therefore observable over the smallest space-time volume. The results for these population assumptions are also shown in Table II and in Fig. 10. The inferred overall rate is shown in Fig. 11. As expected, the population-based rate estimates bracket the one obtained by using the masses of the observed black hole binaries.

TABLE II. Rates of BBH mergers based on populations with masses matching the observed events, and astrophysically motivated mass distributions. Rates inferred from the PyCBC and GstLAL analyses independently as well as combined rates are shown. The table shows median values with 90% credible intervals.

| | $R/(\mathrm{Gpc}^{-3}\ \mathrm{yr}^{-1})$ | | |
|---|---|---|---|
| Mass distribution | PyCBC | GstLAL | Combined |
| | Event based | | |
| GW150914 | $3.2^{+8.3}_{-2.7}$ | $3.6^{+9.1}_{-3.0}$ | $3.4^{+8.8}_{-2.8}$ |
| LVT151012 | $9.2^{+30.3}_{-8.5}$ | $9.2^{+31.4}_{-8.5}$ | $9.1^{+31.0}_{-8.5}$ |
| GW151226 | $35^{+92}_{-29}$ | $37^{+94}_{-31}$ | $36^{+95}_{-30}$ |
| All | $53^{+100}_{-40}$ | $56^{+105}_{-42}$ | $55^{+103}_{-41}$ |
| | Astrophysical | | |
| Flat in log mass | $31^{+43}_{-21}$ | $29^{+43}_{-21}$ | $31^{+42}_{-21}$ |
| Power law (−2.35) | $100^{+136}_{-69}$ | $94^{+137}_{-66}$ | $97^{+135}_{-67}$ |

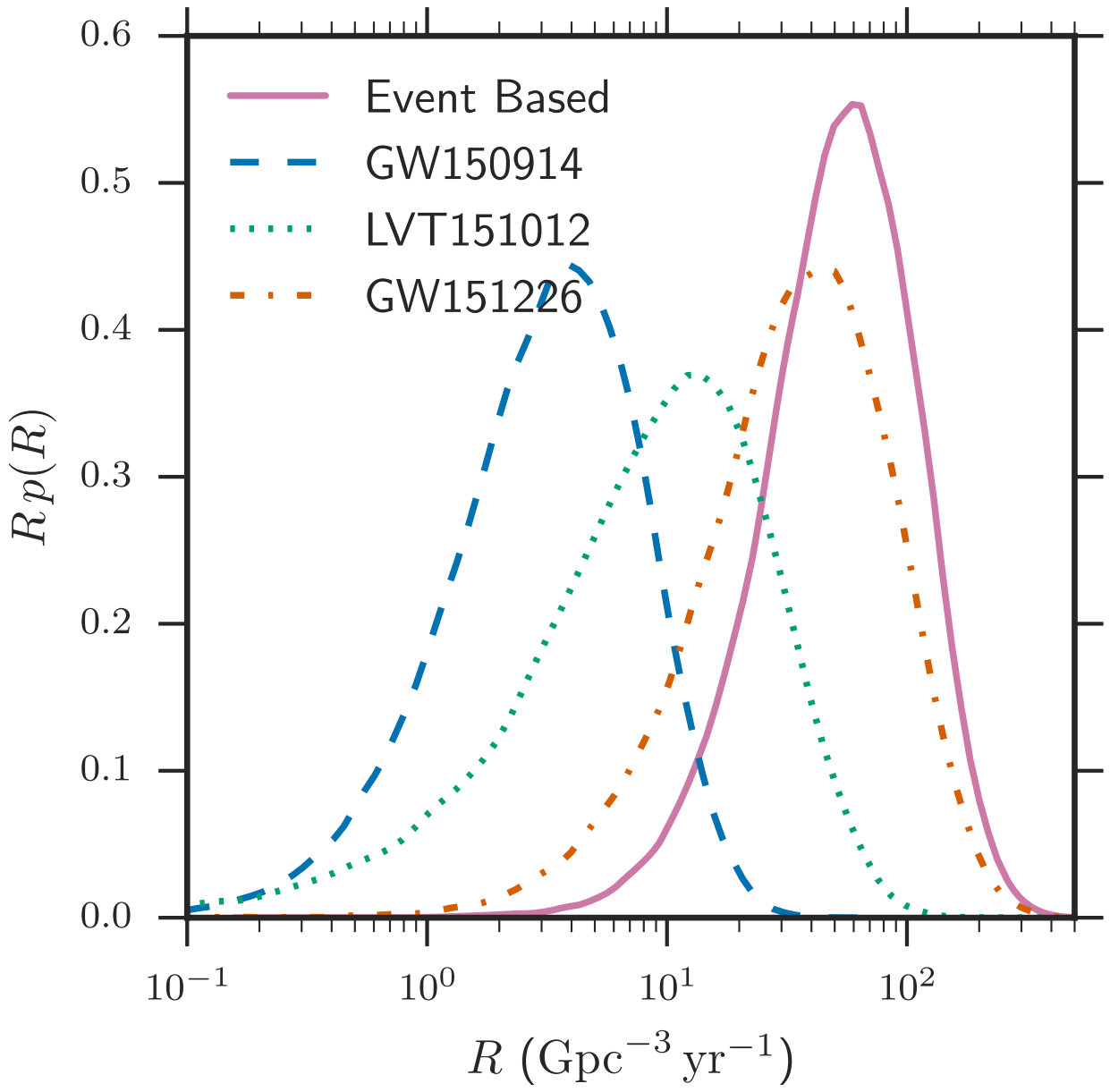

FIG. 10. The posterior density on the rate of GW150914-like BBH, LVT151012-like BBH, and GW151226-like BBH mergers. The event-based rate is the sum of these. The median and 90% credible levels are given in Table II.

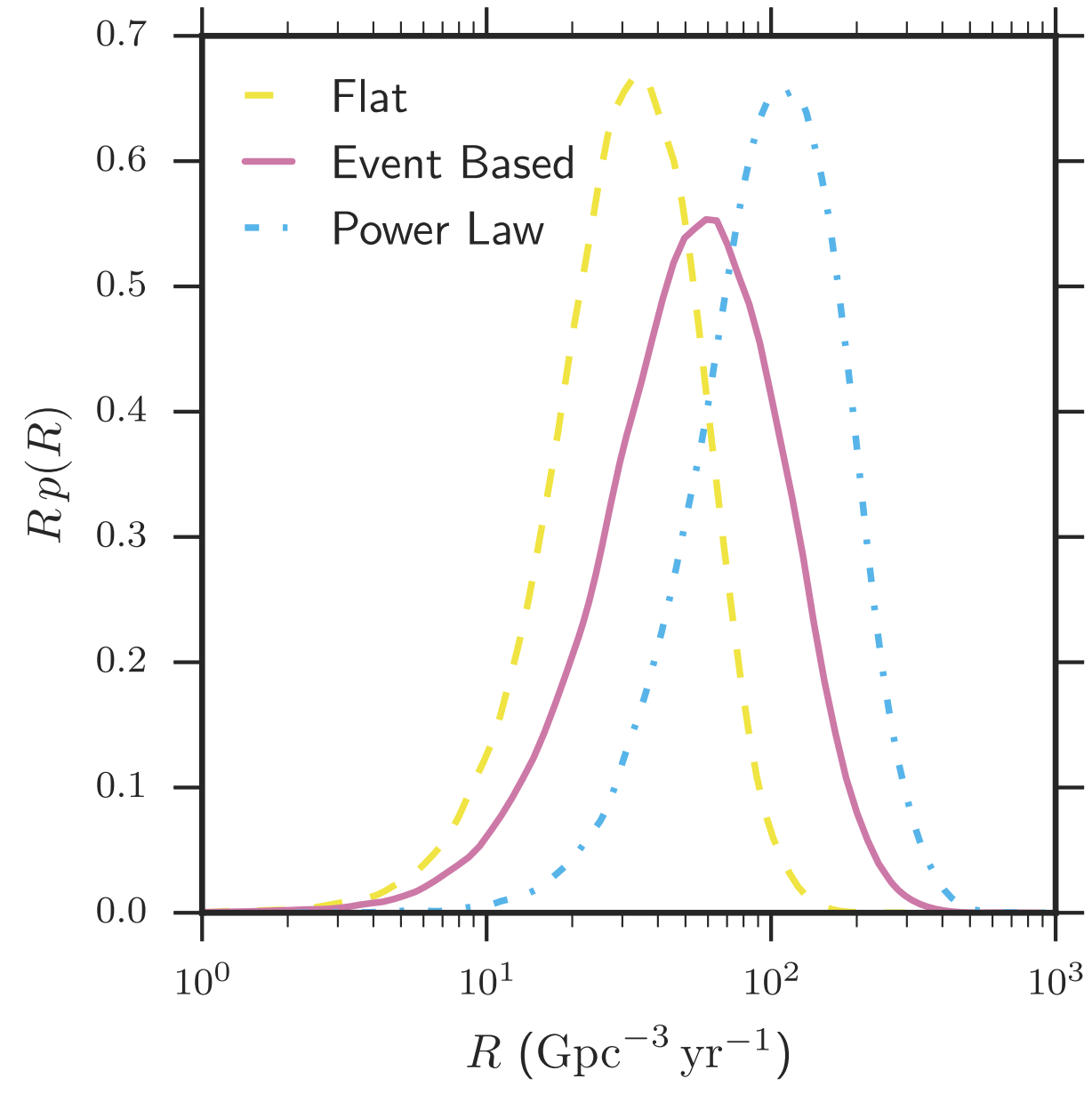

FIG. 11. The posterior density on the rate of BBH mergers. The curves represent the posterior assuming that BBH masses are distributed flat in $\log(m_1) - \log(m_2)$ (Flat), match the properties of the observed events (Event based), or are distributed as a power law in $m_1$ (Power law). The posterior median rates and symmetric 90% symmetric credible intervals are given in Table II.

The inferred rates of BBH mergers are consistent with the results obtained in Refs. [42,155], following the observation of GW150914. The median values of the rates have decreased by approximately a factor of 2, as we now have three likely signals (rather than two) in 3 times as much data. Furthermore, because of the observation of an additional highly significant signal GW151226, the uncertainty in rates has reduced. In particular, the 90% range of allowed rates has been updated to 9–240 $\mathrm{Gpc}^{-3}\,\mathrm{yr}^{-1}$, where the lower limit comes from the flat in log mass population and the upper limit from the power-law population distribution.

With three significant triggers, GW150914, LVT151012, and GW151226, all of astrophysical origin to high probability, we can begin to constrain the mass distribution of coalescing BBHs. Here, we present a simple, parametrized fit to the mass distribution using these triggers; a nonparametric method that can fit general mass distributions will be presented in future work. Our methodology is described more fully in Appendix D.

We assume that the distribution of black hole masses in coalescing binaries follows

$$p(m_1) \propto m_1^{-\alpha}, \tag{7}$$

with $M_{\min} \leq m_2 \leq m_1$ and $m_1 + m_2 \leq 100 \mathrm{M}_\odot$, and a uniform distribution on the secondary mass between $M_{\min} = 5\mathrm{M}_\odot$ and $m_1$. With $\alpha = 2.35$, this mass distribution is the power-law distribution used in our rate estimation. Our choice of $M_{\min}$ is driven by a desire to incorporate nearly all the posterior samples from GW151226 and because there is some evidence from electromagnetic observations for a minimum BH mass near $5\mathrm{M}_\odot$ [82,156] (but see Ref. [84]).

We use a hierarchical analysis [156–159] to infer $\alpha$ from the properties of the three significant events—GW150914, GW151226, and LVT151012—where all three are treated equally and we properly incorporate parameter-estimation uncertainty on the masses of each system. Our inferred posterior on $\alpha$ is shown in Fig. 12. The value $\alpha = 2.35$, corresponding to the power-law mass distribution used above to infer rates, lies near the peak of the posterior, and the median and broad 90% credible interval is

$$\alpha = 2.5^{+1.5}_{-1.6}. \tag{8}$$

It is not surprising that our fit peaks near $\alpha \sim 2.5$ because the observed sample is consistent with a flat distribution and the sensitive space-time volume scales roughly as $M^{15/6}$.

The mass distribution of merging black hole binaries cannot be constrained tightly with such a small number of observations. This power-law fit is sensitive to a number of arbitrary assumptions, including a flat distribution in the mass ratio and a redshift-independent merger rate and mass distribution. Most critically, the fit is sensitive to the choice of the lower-mass cutoff $M_{\min}$: Larger values of $M_{\min}$ lead

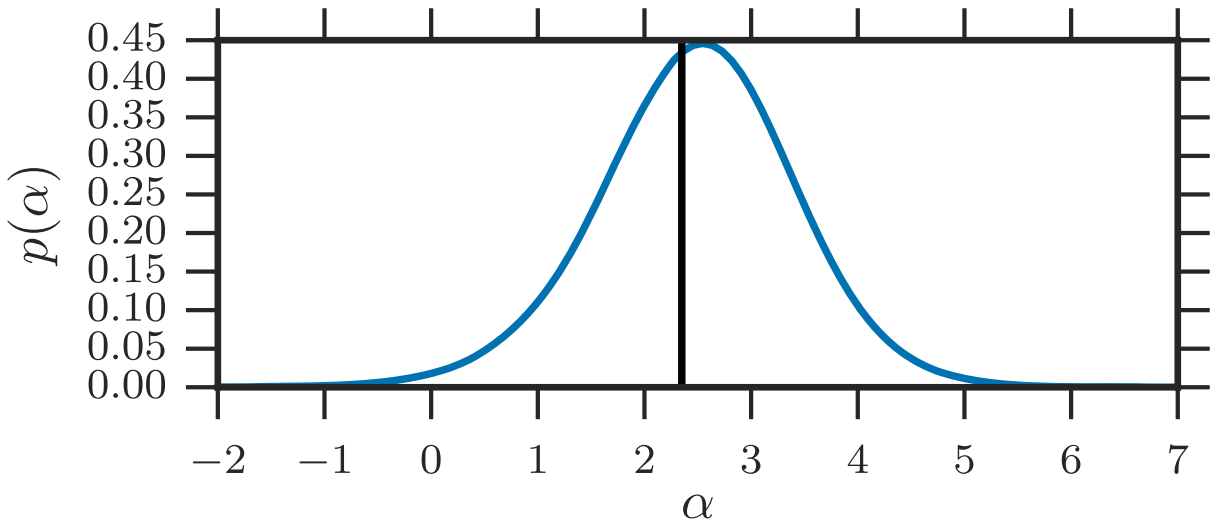

FIG. 12. The posterior distribution for $\alpha$ in Eq. (7) using the inferred masses for our three most significant triggers, GW150914, LVT151012, and GW151226. The vertical line indicates the value of $\alpha = 2.35$ that corresponds to the power-law mass distribution used to infer the rate of BBH coalescence. This value is fully consistent with the posterior, which allows a broad range of possible values with a median and 90% credible interval of $\alpha = 2.5^{+1.5}_{-1.6}$.

to a preference for steeper power laws with indices different by a few.

## VII. ASTROPHYSICAL IMPLICATIONS AND FUTURE PROSPECTS

In Ref. [160], we discussed the astrophysical implications of the first gravitational-wave detection, GW150914, of the merger of two black holes with masses $m_1 = 36.2^{+5.2}_{-3.8} \mathrm{M}_\odot$ and $m_2 = 29.1^{+3.7}_{-4.4} \mathrm{M}_\odot$. We concluded that, while it demonstrated that nature produces BBHs that merge in a Hubble time, it was impossible to determine the formation channel for that event. Possible BBH formation channels include dynamical formation in a dense stellar environment (see, e.g., Refs. [161–165]), possibly assisted by gas drag in galactic nuclear disks [166,167], or isolated binary evolution, either the classical variant via a common-envelope phase (see, e.g., Refs. [168–173]), possibly from population III binaries [174,175], or chemically homogeneous evolution in close tidally locked binaries [176,177]. All of these channels have been shown to be consistent with the GW150914 discovery [178–186].

GW151226 differs from GW150914 primarily in the significantly lower inferred companion masses: $m_1 = 14.2^{+8.3}_{-3.7} \mathrm{M}_\odot$ and $m_2 = 7.5^{+2.3}_{-2.3} \mathrm{M}_\odot$. These masses are similar to the black hole masses measured dynamically in x-ray binaries (for reviews, see Refs. [82,156]). If LVT151012 is of astrophysical origin, its inferred companion masses $m_1 = 23^{+18}_{-6} \mathrm{M}_\odot$ and $m_2 = 13^{+4}_{-5} \mathrm{M}_\odot$ fall between those of GW150914 and GW151226. This result indicates that merging BBHs exist in a broad mass range.

GW151226 and LVT151012 could have formed from lower-mass progenitor stars than GW150914 and/or in higher-metallicity environments in which progenitors lose a greater fraction of their mass to winds. Black holes with such masses can be formed at solar metallicity; see, e.g., Ref. [187]. The low masses of GW151226 are probably inconsistent with the chemically homogeneous evolution scenario, under which higher masses are thought to be required [176,177]. However, the masses are still consistent with both classical isolated binary evolution and dynamical formation.

The broad power-law index range $\alpha = 2.5^{+1.5}_{-1.6}$ inferred from the fit to the merging binary black hole mass distribution attempted in Sec. VI demonstrates the statistical uncertainty associated with extrapolating a distribution from just three events. There are additional systematic uncertainties associated with the power-law model. In particular, while population-synthesis models of binary evolution can be consistent with power-law mass distributions over a range of masses, as in Figs. 8 and 9 of Ref. [188], the power law is likely to be broken over the very broad range between $M_{\mathrm{min}} = 5\ M_\odot$ and a total mass of $100\ M_\odot$. Other formation models may not be consistent with power-law distributions altogether (see, e.g., Ref. [183]). Similar methods have been employed to fit the population of black holes with dynamical mass measurements in x-ray binaries: Reference [156] obtained, for a power-law model, $M_{\mathrm{min}} \sim 5$ and power-law slopes in the range $1.8 \lesssim \alpha \lesssim 5.0$ without accounting for possible selection effects.

Isolated binary evolution is thought to prefer comparable masses, with mass ratios $q < 0.5$ unlikely for the classical scenario [189] and implausible for chemically homogeneous evolution [181]. The dynamical formation channel also prefers comparable masses but allows for more extreme mass ratios; observations of merging binary black holes with extreme mass ratios could therefore point to their dynamical origin. However, the mass ratios of GW151226, $q \geq 0.28$, and LVT151012, $q \geq 0.24$, are not well determined, and $q = 1$ cannot be ruled out for either event. Similarly, spin measurements, which point to a moderate degree of net spin alignment with the orbital angular momentum for GW151226, $\chi_{\mathrm{eff}} = 0.21^{+0.20}_{-0.10}$, cannot be used to distinguish formation channels. On the other hand, a zero effective spin is ruled out for GW151226; the data indicate that at least one of the merging black holes must have been spinning with $a > 0.2$ at the 99% credible level.

The inferred GW151226 merger luminosity distance of $D_{\mathrm{L}} = 440^{+180}_{-190}$Mpc, corresponding to a merger redshift of $z = 0.09^{+0.03}_{-0.04}$, is similar to that of GW150914; in contrast, LVT151012 merged about a factor of 2 further away, at $D_{\mathrm{L}} = 1000^{+500}_{-500}$Mpc, or $z = 0.20^{+0.09}_{-0.09}$. Both are consistent with either a relatively recent formation followed by a prompt merger or formation in the early Universe with a significant time delay between formation and merger.

The BBH merger rate inferred from the full analysis of all O1 triggers, $R = 9$–$240\ \mathrm{Gpc}^{-3}\,\mathrm{yr}^{-1}$, is consistent with the rate inferred from the first 16 days of the O1 run [42]. The full O1 merger rate can be used to update the estimate of the energy density $\Omega_{\mathrm{GW}}$ in the stochastic gravitational-wave background from unresolvable BBH mergers, improving on early results in Ref. [190]. Using the event-based, log-flat, and power-law mass distributions presented in Sec. VI and the corresponding combined rates in Table III, and employing the other "Fiducial" model assumptions from Ref. [190], we obtain 90% credible intervals on $\Omega_{\rm GW}$. The three models agree at frequencies below 100 Hz, where $\Omega_{\rm GW}(f) \sim f^{2/3}$ and which contain more than 99% of the signal-to-noise ratio for stochastic backgrounds, with $\Omega_{\rm GW}(f = 25\ \mathrm{Hz}) \sim 1.2^{+1.9}_{-0.9} \times 10^{-9}$. These predictions do not significantly change the median value of $\Omega_{\rm GW}$ from Ref. [190] while slightly decreasing the range; we still conclude that this background is potentially measurable by the Advanced LIGO/Virgo detectors operating at their projected design sensitivity.

TABLE III. The standard deviations used for the (zero-mean) Gaussian priors on calibration uncertainty for each of the three events. The calibration of each of the two detectors has been independently assessed [47]. These priors set the expected variation for the frequency-dependent spline model used to incorporate the effects of calibration uncertainty [191].

| | Amplitude | | Phase | |
|---|---|---|---|---|
| Event | Hanford | Livingston | Hanford | Livingston |
| GW150914 | 4.8% | 8.2% | 3.2 deg | 4.2 deg |
| LVT151012 | 4.2% | 8.3% | 2.7 deg | 4.3 deg |
| GW151226 | 4.2% | 6.9% | 2.7 deg | 3.6 deg |

Despite the uncertainty in the merger rate, its lower limit can be used to rule out some corners of the parameter space if a single formation channel is assumed for all BBHs. For example, if all merging BBHs arise from dynamical formation in globular clusters, then the lower limit on the merger rate disfavors low-mass clusters [165]. On the other hand, if all merging BBHs arise from isolated binaries evolving via the common-envelope phase, the lower limit on the merger rate disfavors a combination of very-low common envelope binding energy with a high efficiency of common envelope ejection [189] (high values of $\alpha \times \lambda$, as defined in Refs. [192–194]), or very high black hole natal kicks of several hundred km/s [195]. However, since population synthesis studies have typically varied one parameter at a time, individual parameter values cannot be ruled out until the full parameter space is explored (see, e.g., Ref. [196]). Moreover, the parametrizations used in existing models may not even capture the full physical uncertainties (see, e.g., Refs. [197,198]).

It is likely, however, that multiple formation channels are in operation simultaneously, and GW150914, LVT151012, and GW151226 could have been formed through different channels or in different environments. A lower limit on the merger rate cannot be used to rule out evolutionary parameters if multiple channels contribute. Future observations will be required to test whether binaries can be classified into distinct clusters arising from different formation channels [199] or to compare the population to specific evolutionary models [200–203]. Such observations will make it possible to further probe the underlying mass distribution of merging BBHs and the dependence of the merger rate on redshift. Meanwhile, space-borne detectors such as eLISA could observe heavy BBHs several years before merger; multispectrum observations with ground-based and space-borne observatories would aid in measuring binary parameters, including location, and determining the formation channel by measuring the eccentricity at lower frequencies [204–206].

We can use the inferred rates to estimate the number of BBH mergers expected in future observing runs. We make use of the future observing plans laid out in Ref. [132] to predict the expected rate of signals in the second and third advanced LIGO and Virgo observing runs. To do so, we restrict our attention to those signals which will be observed with a false alarm rate smaller than 1/100 yr. In the simulations used to estimate sensitive space-time volumes, 61% of the events above the low threshold used in the PyCBC rates calculation are found with a search false alarm rate lower than one per century. The expected number of observed events will then scale linearly with the sensitive space-time volume $\langle VT \rangle$ of a future search. The improvement in sensitivity in future runs will vary across the frequency band of the detectors and will therefore have a different impact for binaries of different mass. For concreteness, we use a fiducial BBH system with total mass $60 M_\odot$ and mass ratio $q = 1$ [160], to estimate a range of sensitive space-time volumes for future observing runs [207]. The second observing run (O2) is anticipated to begin in late 2016 and last six months, and the third run (O3) is to begin in 2017 and last nine months. We show the predictions for the probability of obtaining $N$ or more high-significance events as a function of $\langle VT \rangle$ (in units of the space-time volume surveyed during O1) in Fig. 13. Current projections for O2 suggest that the sensitivity will be consistent with the lower end of the band indicated in Fig. 13.

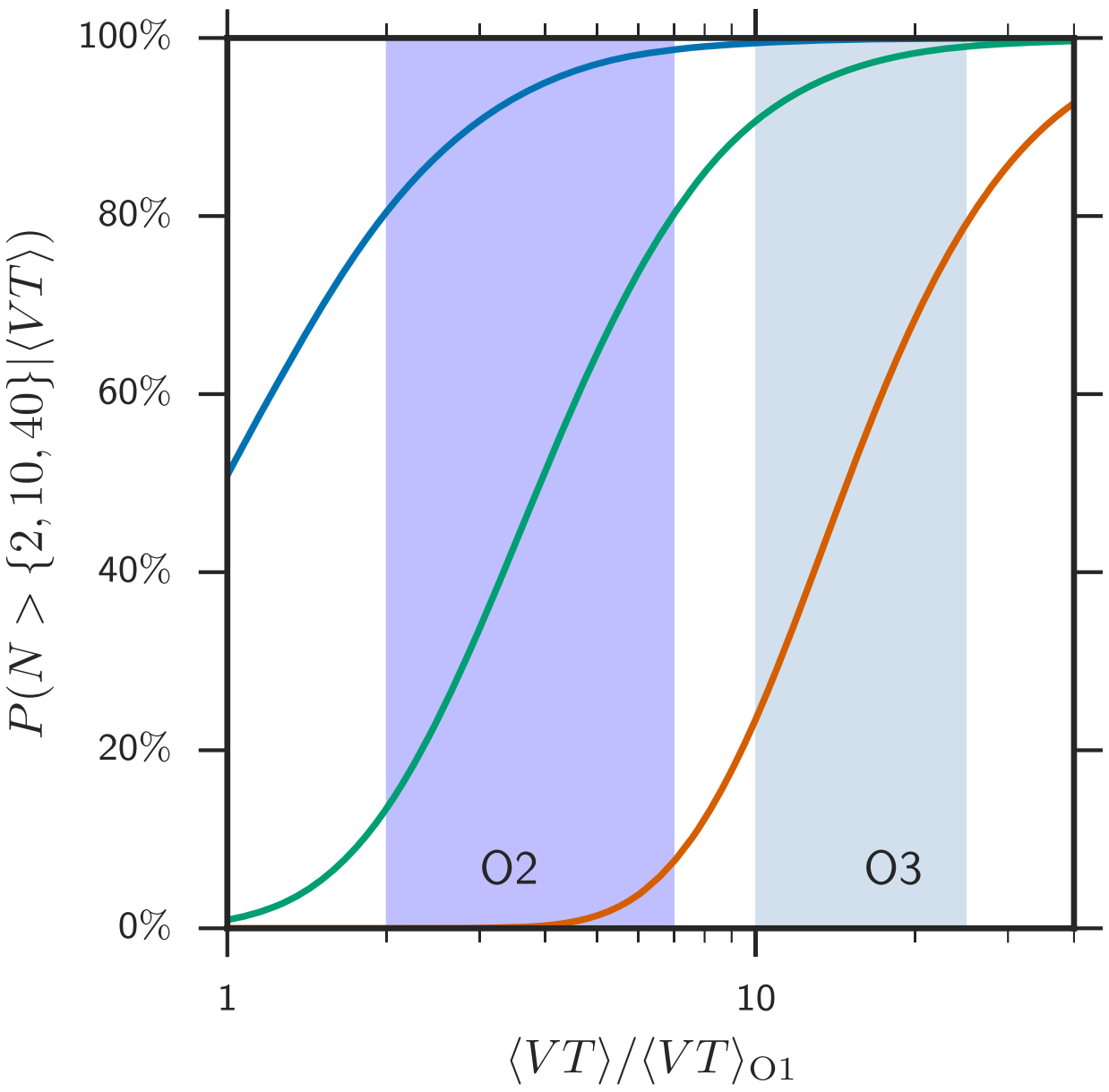

FIG. 13. The probability of observing $N > 2$, $N > 10$, and $N > 40$ highly significant events, as a function of surveyed space-time volume, given the results presented here. The vertical line and bands show, from left to right, the expected sensitive space-time volume for the second (O2) and third (O3) advanced detector observing runs.

## VIII. CONCLUSION

During its first observing run, Advanced LIGO has observed gravitational waves from the coalescence of two stellar-mass BBHs, GW150914 and GW151226, with a third candidate LVT151012 also likely to be a BBH system. Our modeled binary coalescence search detects both GW150914 and GW151226 with a significance of greater than $5.3\sigma$, while LVT151012 is found with a significance of $1.7\sigma$. The component masses of these systems span a range from the heaviest black hole in GW150914 with a mass of $36.2^{+5.2}_{-3.8}\mathrm{M}_\odot$, to $7.5^{+2.3}_{-2.3}\mathrm{M}_\odot$, the lightest black hole of GW151226. The spins of the individual coalescing black holes are weakly constrained, but we can rule out two nonspinning components for GW151226 at the 99% credible level. All our observations are consistent with the predictions of general relativity, and the final black holes formed after merger are all predicted to have high spin values with masses that are larger than any black hole measured in x-ray binaries. The inferred rate of BBH mergers based on our observations is 9–240 $\mathrm{Gpc}^{-3}\,\mathrm{yr}^{-1}$, which gives confidence that future observing runs will observe many more BBHs.

## ACKNOWLEDGMENTS

The authors gratefully acknowledge the support of the United States National Science Foundation (NSF) for the construction and operation of the LIGO Laboratory and Advanced LIGO, as well as the Science and Technology Facilities Council (STFC) of the United Kingdom, the Max-Planck-Society (MPS), and the State of Niedersachsen/Germany for support of the construction of Advanced LIGO and construction and operation of the GEO600 detector. Additional support for Advanced LIGO was provided by the Australian Research Council. The authors gratefully acknowledge the Italian Istituto Nazionale di Fisica Nucleare (INFN), the French Centre National de la Recherche Scientifique (CNRS) and the Foundation for Fundamental Research on Matter supported by the Netherlands Organisation for Scientific Research, for the construction and operation of the Virgo detector and the creation and support of the EGO consortium. The authors also gratefully acknowledge research support from these agencies, as well as by the Council of Scientific and Industrial Research of India; Department of Science and Technology, India; Science & Engineering Research Board (SERB), India; Ministry of Human Resource Development, India; the Spanish Ministerio de Economía y Competitividad; the Conselleria d'Economia i Competitivitat and Conselleria d'Educació; Cultura i Universitats of the Govern de les Illes Balears; the National Science Centre of Poland; the European Commission; the Royal Society; the Scottish Funding Council; the Scottish Universities Physics Alliance; the Hungarian Scientific Research Fund (OTKA); the Lyon Institute of Origins (LIO); the National Research Foundation of Korea; Industry Canada and the Province of Ontario through the Ministry of Economic Development and Innovation; the Natural Science and Engineering Research Council Canada; Canadian Institute for Advanced Research; the Brazilian Ministry of Science, Technology, and Innovation; Fundaćão de Amparo à Pesquisa do Estado de São Paulo (FAPESP); Russian Foundation for Basic Research; the Leverhulme Trust, the Research Corporation; Ministry of Science and Technology (MOST), Taiwan; and the Kavli Foundation. The authors gratefully acknowledge the support of the NSF, STFC, MPS, INFN, CNRS, and the State of Niedersachsen/Germany for provision of computational resources.

## APPENDIX A: SEARCH DESCRIPTION

In this appendix, we give further details of the two analyses, PyCBC and GstLAL, used in the search. Both analyses separately correlate the data from each detector with template waveforms that model the expected signal. The analyses identify candidate events that are detected at both the Hanford and Livingston observatories, consistent with the 10-ms intersite propagation time. Additional signal consistency tests are performed to mitigate the effects of nonstationary transients in the data. Events are assigned a detection-statistic value that ranks their likelihood of being a gravitational-wave signal. This detection statistic is compared to the estimated detector noise background to determine, for each candidate event, the probability that detector noise would give rise to at least one equally significant event.

The choice of parameters for the templates depends on the shape of the power spectrum of the detector noise. The average noise power spectral density of the LIGO detectors was measured over the period September 12 to September 26, 2015. The harmonic mean of these noise spectra from the two detectors was used to place a single template bank that was employed for the duration of the search [3].

The matched-filter SNR $\rho$ for each template waveform and each detector's data as a function of time is calculated according to [11,208]

$$\rho^2(t) \equiv [\langle s|h_c\rangle^2(t) + \langle s|h_s\rangle^2(t)], \tag{A1}$$

where the correlation is defined by

$$\langle s|h\rangle(t) \equiv 4\mathrm{Re}\int_0^\infty \frac{\tilde{s}(f)\tilde{h}^*(f)}{S_n(f)} e^{2\pi i f t} \mathrm{d}f, \quad \text{(A2)}$$

$h_c$ and $h_s$ are the normalized orthogonal sine and cosine parts of the template, and $\tilde{a}(f)$ is used to denote the Fourier transform of the time domain quantity $a(t)$. Here, $S_n(f)$ denotes the one-sided average power spectral density of the detector noise. The waveform components $h_c$ and $h_s$ are normalized such that the expected value of $\langle s|h_{s,c}\rangle^2(t)$ in stationary, Gaussian noise is unity [95]. The analyses identify times when the matched-filter SNR achieves a local maximum and store each of these as a trigger. The analyses search only stretches of data longer than a minimum duration, to ensure that the detectors are operating stably. The choice is different in the two analyses and reduces the available data of 48.6 days to 46.1 days for the PyCBC analysis and 48.3 days for the GstLAL analysis.

To suppress large SNR values caused by non-Gaussian detector noise, the analyses perform additional tests to quantify the agreement between the data and the template. These tests are different in the two analyses and are discussed in their respective subsections below. Both analyses enforce coincidence between detectors by selecting trigger pairs that occur within a 15-ms window and come from the same template. The 15-ms window is determined by the 10-ms intersite propagation time plus 5 ms for uncertainty in accurately determining the measured arrival time of weak signals. A detection statistic for each coincident event is derived as a function of the SNR observed in each detector, the value of the signal consistency tests, and details of the template.

The significance of a candidate event is determined by comparing it to the search background. From this, we are able to determine the rate at which detector noise produces events with a detection-statistic value equal to or higher than the candidate event (the FAR). Estimating this background is challenging for two reasons: First, the detector noise is nonstationary and non-Gaussian; therefore, its properties must be empirically determined. Second, it is not possible to shield the detector from gravitational waves to directly measure a signal-free background. The specific procedure used to estimate the background is different for the two analyses, as described in detail below.

The results of the independent analyses are two separate lists of candidate events, with each candidate event assigned a p-value and FAR. Candidate events with low FARs are identified as possible gravitational-wave signals for further investigation.

### 1. PyCBC analysis

The PyCBC analysis is described in detail in Refs. [2–4], and the configuration used to analyze the first 16 days of O1 data, containing GW150914, is described in Ref. [44]. Following the observation of GW150914, some improvements were made to the analysis, as we better understood the Advanced LIGO data. All changes were tested and tuned only on background data, prior to being incorporated into the analysis. These changes do not affect the significance bound of GW150914. Consequently, we chose to present the full results, on the final calibrated data, using the improved analysis. Here, we provide a brief overview of the analysis, including details of changes made following the discovery of GW150914.

In the PyCBC analysis, a trigger is stored when the maximum of the SNR time series is above the threshold of 5.5 (chosen as a compromise between a manageable trigger rate and assurance that no real event will be missed), with a maximum of one trigger stored in a 1-s window (reduced from 4 s in the previous analysis). A $\chi^2$ statistic is computed to distinguish between astrophysical signals and noise transients. This result tests whether the signal power in a number of nonoverlapping frequency bands is consistent with that expected from the waveform template [14]. The $\chi^2$ test is written explicitly as

$$\chi_r^2 = \frac{p}{2p-2}\sum_{i=1}^{p}\left(\rho_i - \frac{\rho}{p}\right)^2, \quad \text{(A3)}$$

where $p$ denotes the number of frequency bands—constructed such that the expected signal power in each band is equal—and $\rho_i$ is the matched-filter SNR in the $i$th frequency band. For data containing only Gaussian noise, or Gaussian noise and a signal exactly matching the template waveform, the expected value of this statistic will be 1. For data containing non-Gaussian artefacts, or a signal not matching well with the template waveform, this value will be elevated. Each trigger is then ranked according to a combination of the SNR and the $\chi^2$ test, namely,

$$\hat{\rho} = \begin{cases} \rho[(1+(\chi_r^2)^3)/2]^{-1/6} & \text{if } \chi_r^2 > 1 \\ \rho & \text{if } \chi_r^2 \leq 1. \end{cases} \quad \text{(A4)}$$

The number of frequency bands $p$ used to compute the $\chi^2$ signal-based veto [14] was optimized using data from the first month of O1. An improved background rejection was found when adopting the following, template-dependent expression for the number of $\chi^2$ bands,

$$p = 1.75 \times \left(\frac{f_{\mathrm{peak}}}{1\ \mathrm{Hz}} - 60\right)^{1/2}, \quad \text{(A5)}$$

where $f_{\mathrm{peak}}$ is the frequency corresponding to the maximum amplitude of the template waveform using the models described in Ref. [8], and $p$ is rounded to the nearest integer. This choice was adopted for the full O1 analysis presented here, where all waveforms have peak frequencies greater than 60 Hz.

Loud and short instrumental transients are identified and excised from the data, as part of the data conditioning prior to SNR computation. In this analysis, we compute a whitened time series of the strain data and compare the magnitude of each sample against a threshold value of 100. Samples above threshold and within a time window of $\pm 0.5$ s are clustered together, and a gating window is placed at the time of the loudest sample in the cluster [209]. The threshold value of 100 is chosen to be much larger than the typical value of the magnitude in Gaussian noise and also larger than the value expected from any gravitational-wave signal from binaries at astrophysical distances and with intrinsic parameters within our search space.

Coincident triggers are formed when a trigger exists in both observatories, associated with the same template waveform and with arrival times within 15 ms. Each coincidence is ranked with a network statistic $\hat{\rho}_c$, defined as the quadrature sum of the $\hat{\rho}$ in each observatory. The rate of background events, as a function of network statistic, is estimated from the data themselves by repeating the analysis after artificially time-shifting the triggers from one detector relative to the other. Time shifts in multiples of 100 ms are performed, leading to a total of $T_b = 5.0 \times 10^6$ years of background time analyzed.

The distribution of background noise events over $\hat{\rho}_c$ can vary strongly as a function of the template waveform; to account for this variation, the parameter space is divided into a number of regions which are treated as independent searches [44]. Each coincident trigger is assigned a FAR based on the background distribution in the region containing the coincidence and incorporating a trial factor equal to the number of regions. Studies of the background distribution as a function of the template parameters, and a reduced rate of noise events in O1 data, compared to the engineering run data previously used in tuning the search configuration [44], motivated a redefinition of the regions used to divide the search space. In the current analysis, we split the parameter space into three regions, defined by (i) $\mathcal{M} < 1.74 \mathrm{M}_\odot$, (ii) $\mathcal{M} \geq 1.74 \mathrm{M}_\odot$ and $f_{\mathrm{peak}} \geq 100$ Hz, and (iii) $\mathcal{M} \geq 1.74 \mathrm{M}_\odot$ and $f_{\mathrm{peak}} < 100$ Hz. In the GW150914 analysis, the boundary between regions (ii) and (iii) was set at 220 Hz. By reducing this frequency, we significantly reduce the number of templates assigned to region (iii), which is dominated by short templates that are most affected by noise transients. The frequency at peak amplitude of the best-matching template for GW150914 is $f_{\mathrm{peak}} = 144$ Hz. With the tuning used for the original result, this placed it in noise-background class (iii) of the PyCBC analysis [44]. However, with the improved O1 tuning, which changed the boundaries of the noise-background classes, this event is in noise-background class (ii). In Fig. 3, we plot the background only from class (ii), while the quoted significances take into account a trial factor of three because of the three noise-background classes.

### 2. GstLAL analysis

The GstLAL [210] analysis method is a low-latency, multidetector matched-filtering search for gravitational waves emitted by the coalescence of compact objects. The analysis exploits time-domain operations [5] that give it a latency of seconds after the acquisition of gravitational-wave data. This allows the GstLAL analysis to run in both low-latency mode to provide rapid identification of signals and in off-line mode on data that have been conditioned with data-quality vetoes [13]. The results presented here are for the off-line mode. No changes were made to the GstLAL analysis relative to the results presented in Ref. [44].

For the off-line analysis, the data $s(t)$ are partitioned into chunks, and along with the templates $h(t)$, the data $s(t)$ are then whitened in the frequency domain. The analysis splits the template bank into sub-banks containing waveforms that have morphological similarities. The templates are binned in a two-dimensional space by effective spin parameter $\chi_{\mathrm{eff}}$ and chirp mass $\mathcal{M}$, as these parameters can be used to effectively describe a binary system in which the spins are aligned with the binary's orbital angular momentum. Templates are allowed to overlap in adjacent bins to mitigate boundary effects, although no redundant waveforms are filtered.

An orthonormal basis of filters $\hat{h}(t)$ is then constructed using singular value decomposition [5]. This basis is significantly smaller than the number of input waveforms and allows for a significant reduction in the time-domain filtering cost. The set of filters $\hat{h}(t)$ in each bin is convolved with the whitened data, producing a time series; the matched-filter SNR time series $\rho$ for each template can then be constructed using linear combinations of the convolution time series. A trigger is stored when the maximum of the SNR time series crosses a predetermined threshold of 4. A maximum of one coincident trigger per template is stored in each second.

A signal consistency test is performed by comparing the SNR time series of data to the SNR time series *expected* from a real signal using the autocorrelation function of the template at its time of peak amplitude, $R(t)$. A consistency test value $\xi^2_{\mathrm{ac}}$ is determined for each trigger using the SNR time series $\rho(t)$, the peak SNR $\rho_p$, and the autocorrelation function $R(t)$ in some window of time $\delta t$ (corresponding to $\rho_p$) around the trigger:

$$\xi^2_{\mathrm{ac}} = \frac{1}{\mu} \int_{t_p+\delta t}^{t_p-\delta t} dt |\rho(t) - \rho_p R(t)|^2, \qquad \text{(A6)}$$

where the factor $\mu$ ensures that a well-fit signal has an expectation value of 1 [44]. The window $\delta t$ is a tunable parameter that has been chosen based on Monte Carlo simulations in real data and finding the value that (on average) best rejects glitches.

Triggers that survive consistency checks are assigned a ranking based upon their SNR, $\xi^2_{\mathrm{ac}}$ value, and the instantaneous horizon distance values at each detector, $\{D_{\mathrm{H1}}, D_{\mathrm{L1}}\}$, which encode the detector sensitivity [15,211].

A likelihood ratio is constructed to rank candidate events by the ratio of the probability of observing matched-filter SNR and $\xi^2_{\rm ac}$ from signals, h, versus obtaining the same parameters from noise, n. The templates have already been grouped into regions that contain high overlap, so it is likely that templates within each group will respond similarly to noise; in fact, the template group itself is used as a parameter in the likelihood ratio to qualitatively establish how different regions of the parameter space are affected by noise. The likelihood ratio can thus be written as

$$\mathcal{L} = \frac{p(\mathbf{x}_{\rm H}, \mathbf{x}_{\rm L}, D_{\rm H}, D_{\rm L}|\theta_i, {\rm h})}{p(\mathbf{x}_{\rm H}|\theta_i, {\rm n})p(\mathbf{x}_{\rm L}|\theta_i, {\rm n})}, \qquad (A7)$$

where $\mathbf{x}_d = \{\rho_d, \xi^2{}_d\}$ are the matched-filter SNR and $\xi^2_{\rm ac}$ in each detector, $\theta_i$ corresponds to the template group, and $D_d$ is the horizon distance of the given detector at the time of the trigger. The signal distribution in the numerator is calculated using an astrophysical model of signals distributed isotropically in the nearby universe. The denominator is calculated under the assumption that the noise in each detector is independent. It can then be calculated from the distribution of triggers in each template bin observed in each detector. In the case where multiple high-likelihood events are produced at the same time, a clustering process is used to remove events with lower likelihoods within a 4-s window so that only the event with the highest likelihood is retained.

In a typical search, the majority of events found in coincidence correspond to noise and not an actual signal. To accurately distill signals from the data, the p-value at the value of $\mathcal{L}$ for each event is ascertained; the p-value describes the probability of observing the event's $\mathcal{L}$ or greater in noise alone. The GstLAL method determines the p-value by taking the probability density functions of parameters in Eq. (A7) obtained from triggers that are noiselike in nature [212].

## APPENDIX B: PARAMETER-ESTIMATION DESCRIPTION

To extract information from the signal, we perform a coherent Bayesian analysis of the data from the two instruments using LALInference [48,213]. The properties of the source leave imprints on the signal from which we can infer their values [39]. We match the measured strain to model waveforms and use the agreement to define probability distributions for the parameters that describe the signal. A summary of results for the three events is given in Table IV.

The result of our analysis is the posterior probability distribution for parameters describing the source. The posterior is computed from Bayes' theorem [216,217]: It is proportional to the product of the likelihood of the data given the parameters and the prior for the parameters. The likelihood is calculated using a noise-weighted inner product between the data and the model waveform [95]. This depends upon the waveform and the noise spectral density at the time of event, and both could potentially be sources of systematic error. We incorporate the effects of uncertainty in the detectors' calibration using a frequency-dependent model [191]. The posterior probability density is mapped out using stochastic sampling algorithms, and our parameter estimates are constructed from the distribution of samples.

The analysis makes use of two inspiral-merger-ringdown waveform models, a reduced-order model of the double aligned spin EOB waveform used for the detection analyses, which we refer to as EOBNR [8,9], and an effective precessing spin model, which we refer to as IMRPhenom [36–38,218]. For all events, the results from the EOBNR and IMRPhenom waveforms are similar. An analysis using a fully precessing EOBNR waveform [219], as done in Ref. [40], will be reported in the future; this analysis is currently too computationally expensive for results to be presented now.

To compare how well the different waveform models match the data, we use the Bayes factor $\mathcal{B}_{\rm s/n}$ and the deviance information criterion (DIC). The Bayes factor is the ratio of the evidence (the marginalized likelihood) for a coherent signal hypothesis to that for Gaussian noise [215]. A larger Bayes factor indicates that there is more support for the signal model [220]. The DIC is a measure of the goodness of fit of a model, defined as an average log likelihood plus a penalty factor for higher dimensional models [221–223]. A smaller value of the DIC indicates a greater expectation that the model would predict data similar to that being analyzed, and hence that it is a better fit. The values for both quantities are similar for all three events. The data do not allow us to conclusively prefer one waveform model over the other; therefore, in the column titled Overall of Table IV, results are constructed by averaging the two, marginalizing over our choice of waveform.

Inaccuracies in the waveform models could be a source of systematic error in the parameter estimates [224–226]. However, an alternative analysis of GW150914 using a set of waveforms from numerical-relativity simulations yielded results consistent with those using the EOBNR and IMRPhenom approximants [227]. For our results, we use the difference between results from the two waveform models as a proxy for the theoretical error from waveform modeling, although some known physics such as higher modes and eccentricity are missing from both of these waveform families. For each parameter, we quote systematic errors on the boundaries of the 90% credible intervals; this is the 90% range of a normal distribution estimated from the variance of results from the different models [39]. For parameters with bounded ranges, like the spins or mass ratio, the normal distributions should be truncated, but for simplicity, we still quote the 90% range of the uncut distributions. More sophisticated means of incorporating waveform uncertainty into the analysis, such as Gaussian

TABLE IV. Parameters that characterize GW150914, GW151226, and LVT151012. For model parameters, we report the median value with the range of the symmetric 90% credible interval [214]; we also quote selected 90% credible bounds. For the logarithm of the Bayes factor for a signal compared to Gaussian noise, we report the mean and its 90% standard error from four parallel runs with a nested sampling algorithm [215], and for the deviance information criterion, we report the mean and its 90% standard error from a Markov-chain Monte Carlo and a nested sampling run. The source redshift and source-frame masses assume standard cosmology [18]. Results are given for spin-aligned EOBNR and precessing IMRPhenom waveform models. The "Overall" results are computed by averaging the posteriors for the two models. For the overall results, we quote both the 90% credible interval or bound and an estimate for the 90% range of systematic error on this determined from the variance between waveform models. Further explanations of the parameters are given in Ref. [39].

| | GW150914 | | | GW151226 | | | LVT151012 | | |
|---|---|---|---|---|---|---|---|---|---|
| | EOBNR | IMRPhenom | Overall | EOBNR | IMRPhenom | Overall | EOBNR | IMRPhenom | Overall |
| Detector frame | | | | | | | | | |
| Total mass $M/\mathrm{M}_\odot$ | $71.0^{+4.6}_{-4.0}$ | $71.2^{+3.5}_{-3.2}$ | $71.1^{+4.1\pm0.7}_{-3.6\pm0.8}$ | $23.6^{+8.0}_{-1.3}$ | $23.8^{+5.1}_{-1.5}$ | $23.7^{+6.5\pm2.2}_{-1.4\pm0.1}$ | $45^{+17}_{-4}$ | $44^{+12}_{-3}$ | $44^{+16\pm5}_{-3\pm0}$ |
| Chirp mass $\mathcal{M}/\mathrm{M}_\odot$ | $30.4^{+2.3}_{-1.6}$ | $30.7^{+1.5}_{-1.5}$ | $30.6^{+1.9\pm0.3}_{-1.6\pm0.4}$ | $9.71^{+0.08}_{-0.07}$ | $9.72^{+0.06}_{-0.06}$ | $9.72^{+0.07\pm0.01}_{-0.06\pm0.01}$ | $18.1^{+1.3}_{-0.9}$ | $18.1^{+0.8}_{-0.8}$ | $18.1^{+1.0\pm0.5}_{-0.8\pm0.1}$ |
| Primary mass $m_1/\mathrm{M}_\odot$ | $40.2^{+5.2}_{-4.8}$ | $38.5^{+5.4}_{-3.3}$ | $39.4^{+5.4\pm1.3}_{-4.1\pm0.2}$ | $15.3^{+10.8}_{-3.8}$ | $15.8^{+7.2}_{-4.0}$ | $15.6^{+9.0\pm2.6}_{-4.0\pm0.2}$ | $29^{+23}_{-8}$ | $27^{+19}_{-6}$ | $28^{+21\pm5}_{-7\pm0}$ |
| Secondary mass $m_2/\mathrm{M}_\odot$ | $30.6^{+5.1}_{-4.2}$ | $32.7^{+3.1}_{-4.9}$ | $31.7^{+4.0\pm0.1}_{-4.9\pm1.2}$ | $8.3^{+2.5}_{-2.9}$ | $8.1^{+2.5}_{-2.1}$ | $8.2^{+2.6\pm0.2}_{-2.5\pm0.5}$ | $15^{+5}_{-6}$ | $16^{+4}_{-6}$ | $16^{+5\pm0}_{-6\pm1}$ |
| Final mass $M_\mathrm{f}/\mathrm{M}_\odot$ | $67.8^{+4.0}_{-3.6}$ | $67.9^{+3.2}_{-2.9}$ | $67.8^{+3.7\pm0.6}_{-3.3\pm0.7}$ | $22.5^{+8.2}_{-1.4}$ | $22.8^{+5.3}_{-1.6}$ | $22.6^{+6.7\pm2.2}_{-1.5\pm0.1}$ | $43^{+17}_{-4}$ | $42^{+13}_{-2}$ | $42^{+16\pm5}_{-3\pm0}$ |
| Source frame | | | | | | | | | |
| Total mass $M^\mathrm{source}/\mathrm{M}_\odot$ | $65.5^{+4.4}_{-3.9}$ | $65.1^{+3.6}_{-3.1}$ | $65.3^{+4.1\pm1.0}_{-3.4\pm0.3}$ | $21.6^{+7.4}_{-1.6}$ | $21.9^{+4.7}_{-1.7}$ | $21.8^{+5.9\pm2.0}_{-1.7\pm0.1}$ | $38^{+15}_{-5}$ | $37^{+11}_{-4}$ | $37^{+13\pm4}_{-4\pm0}$ |
| Chirp mass $\mathcal{M}^\mathrm{source}/\mathrm{M}_\odot$ | $28.1^{+2.1}_{-1.6}$ | $28.1^{+1.6}_{-1.4}$ | $28.1^{+1.8\pm0.4}_{-1.5\pm0.2}$ | $8.87^{+0.35}_{-0.28}$ | $8.90^{+0.31}_{-0.27}$ | $8.88^{+0.33\pm0.01}_{-0.28\pm0.04}$ | $15.2^{+1.5}_{-1.1}$ | $15.0^{+1.3}_{-1.0}$ | $15.1^{+1.4\pm0.3}_{-1.1\pm0.0}$ |
| Primary mass $m_1^\mathrm{source}/\mathrm{M}_\odot$ | $37.0^{+4.9}_{-4.4}$ | $35.3^{+5.1}_{-3.1}$ | $36.2^{+5.2\pm1.4}_{-3.8\pm0.4}$ | $14.0^{+10.0}_{-3.5}$ | $14.5^{+6.6}_{-3.7}$ | $14.2^{+8.3\pm2.4}_{-3.7\pm0.2}$ | $24^{+19}_{-7}$ | $23^{+16}_{-5}$ | $23^{+18\pm5}_{-6\pm0}$ |
| Secondary mass $m_2^\mathrm{source}/\mathrm{M}_\odot$ | $28.3^{+4.6}_{-3.9}$ | $29.9^{+3.0}_{-4.5}$ | $29.1^{+3.7\pm0.0}_{-4.4\pm0.9}$ | $7.5^{+2.3}_{-2.6}$ | $7.4^{+2.3}_{-2.0}$ | $7.5^{+2.3\pm0.2}_{-2.3\pm0.4}$ | $13^{+4}_{-5}$ | $14^{+4}_{-5}$ | $13^{+4\pm0}_{-5\pm0}$ |
| Final mass $M_\mathrm{f}^\mathrm{source}/\mathrm{M}_\odot$ | $62.5^{+3.9}_{-3.5}$ | $62.1^{+3.3}_{-2.8}$ | $62.3^{+3.7\pm0.9}_{-3.1\pm0.2}$ | $20.6^{+7.6}_{-1.6}$ | $20.9^{+4.8}_{-1.8}$ | $20.8^{+6.1\pm2.0}_{-1.7\pm0.1}$ | $36^{+15}_{-4}$ | $35^{+11}_{-3}$ | $35^{+14\pm4}_{-4\pm0}$ |
| Energy radiated $E_\mathrm{rad}/(\mathrm{M}_\odot c^2)$ | $2.98^{+0.55}_{-0.40}$ | $3.02^{+0.36}_{-0.36}$ | $3.00^{+0.47\pm0.13}_{-0.39\pm0.07}$ | $1.02^{+0.09}_{-0.24}$ | $0.99^{+0.11}_{-0.17}$ | $1.00^{+0.10\pm0.01}_{-0.20\pm0.03}$ | $1.48^{+0.39}_{-0.41}$ | $1.51^{+0.29}_{-0.44}$ | $1.50^{+0.33\pm0.05}_{-0.43\pm0.01}$ |
| Mass ratio $q$ | $0.77^{+0.20}_{-0.18}$ | $0.85^{+0.13}_{-0.21}$ | $0.81^{+0.17\pm0.02}_{-0.20\pm0.04}$ | $0.54^{+0.40}_{-0.33}$ | $0.51^{+0.39}_{-0.25}$ | $0.52^{+0.40\pm0.03}_{-0.29\pm0.04}$ | $0.53^{+0.42}_{-0.34}$ | $0.60^{+0.35}_{-0.37}$ | $0.57^{+0.38\pm0.01}_{-0.37\pm0.04}$ |
| Effective inspiral spin $\chi_\mathrm{eff}$ | $-0.08^{+0.17}_{-0.14}$ | $-0.05^{+0.11}_{-0.12}$ | $-0.06^{+0.14\pm0.02}_{-0.14\pm0.04}$ | $0.21^{+0.24}_{-0.11}$ | $0.22^{+0.15}_{-0.08}$ | $0.21^{+0.20\pm0.07}_{-0.10\pm0.03}$ | $0.06^{+0.31}_{-0.24}$ | $0.01^{+0.26}_{-0.17}$ | $0.03^{+0.31\pm0.08}_{-0.20\pm0.02}$ |
| Primary spin magnitude $a_1$ | $0.33^{+0.39}_{-0.29}$ | $0.30^{+0.54}_{-0.27}$ | $0.32^{+0.47\pm0.10}_{-0.29\pm0.01}$ | $0.42^{+0.35}_{-0.37}$ | $0.55^{+0.35}_{-0.42}$ | $0.49^{+0.37\pm0.11}_{-0.42\pm0.07}$ | $0.31^{+0.46}_{-0.27}$ | $0.31^{+0.50}_{-0.28}$ | $0.31^{+0.48\pm0.03}_{-0.28\pm0.00}$ |
| Secondary spin magnitude $a_2$ | $0.62^{+0.35}_{-0.54}$ | $0.36^{+0.53}_{-0.33}$ | $0.48^{+0.47\pm0.08}_{-0.43\pm0.03}$ | $0.51^{+0.44}_{-0.46}$ | $0.52^{+0.42}_{-0.47}$ | $0.52^{+0.43\pm0.01}_{-0.47\pm0.00}$ | $0.49^{+0.45}_{-0.44}$ | $0.42^{+0.50}_{-0.38}$ | $0.45^{+0.48\pm0.02}_{-0.41\pm0.01}$ |
| Final spin $a_\mathrm{f}$ | $0.68^{+0.05}_{-0.07}$ | $0.68^{+0.06}_{-0.05}$ | $0.68^{+0.05\pm0.01}_{-0.06\pm0.02}$ | $0.73^{+0.05}_{-0.06}$ | $0.75^{+0.07}_{-0.05}$ | $0.74^{+0.06\pm0.03}_{-0.06\pm0.03}$ | $0.65^{+0.09}_{-0.10}$ | $0.66^{+0.08}_{-0.10}$ | $0.66^{+0.09\pm0.00}_{-0.10\pm0.02}$ |
| Luminosity distance $D_\mathrm{L}/\mathrm{Mpc}$ | $400^{+160}_{-180}$ | $440^{+140}_{-170}$ | $420^{+150\pm20}_{-180\pm40}$ | $450^{+180}_{-210}$ | $440^{+170}_{-180}$ | $440^{+180\pm20}_{-190\pm10}$ | $1000^{+540}_{-490}$ | $1030^{+480}_{-480}$ | $1020^{+500\pm20}_{-490\pm40}$ |
| Source redshift $z$ | $0.086^{+0.031}_{-0.036}$ | $0.094^{+0.027}_{-0.034}$ | $0.090^{+0.029\pm0.003}_{-0.036\pm0.008}$ | $0.096^{+0.035}_{-0.042}$ | $0.092^{+0.033}_{-0.037}$ | $0.094^{+0.035\pm0.004}_{-0.039\pm0.001}$ | $0.198^{+0.091}_{-0.092}$ | $0.204^{+0.082}_{-0.088}$ | $0.201^{+0.086\pm0.003}_{-0.091\pm0.008}$ |
| Upper bound | | | | | | | | | |
| Primary spin magnitude $a_1$ | 0.62 | 0.73 | $0.67 \pm 0.09$ | 0.68 | 0.83 | $0.77 \pm 0.12$ | 0.64 | 0.69 | $0.67 \pm 0.04$ |
| Secondary spin magnitude $a_2$ | 0.93 | 0.80 | $0.90 \pm 0.12$ | 0.90 | 0.89 | $0.90 \pm 0.01$ | 0.89 | 0.85 | $0.87 \pm 0.04$ |
| Lower bound | | | | | | | | | |
| Mass ratio $q$ | 0.62 | 0.68 | $0.65 \pm 0.05$ | 0.25 | 0.30 | $0.28 \pm 0.04$ | 0.22 | 0.28 | $0.24 \pm 0.05$ |
| Log Bayes factor $\ln \mathcal{B}_\mathrm{s/n}$ | $287.7 \pm 0.1$ | $289.8 \pm 0.3$ | ... | $59.5 \pm 0.1$ | $60.2 \pm 0.2$ | ... | $22.8 \pm 0.2$ | $23.0 \pm 0.1$ | ... |
| Information criterion DIC | $32977.2 \pm 0.3$ | $32973.1 \pm 0.1$ | ... | $34296.4 \pm 0.2$ | $34295.1 \pm 0.1$ | ... | $94695.8 \pm 0.0$ | $94692.9 \pm 0.0$ | ... |

process regression [228], may be used in the future. For all three events, we find that the theoretical uncertainty from waveform modeling is less significant than statistical uncertainty from the finite SNR of the events.

The calibration error is modeled using a cubic spline polynomial [39,191], and we marginalize over uncertainty in the calibration. Each analysis assumes a prior for the calibration uncertainty, which is specific for each detector at the time of that signal. Standard deviations of the prior distributions for the amplitude and phase uncertainty are given in Table III. The updated calibration uncertainty is better than the original 10% in amplitude and 10 deg in phase [47] used for the first results.

Aside from the difference in calibration, the analysis of GW150914 follows the specification in Ref. [39]. We analyze 8 s of data centered on the time reported by the detection analyses [44], using the frequency range between 20 Hz and 1024 Hz. The time interval is set by the in-band duration of waveforms in the prior mass range. We assume uninformative prior distributions for the parameters (uniform distributions for the time and phase of coalescence, uniform distribution of sources in volume, isotropic orientations for the binary and the two spins, uniform distribution of spin magnitudes, and uniform distribution of component masses $m_{1,2} \in [10, 80] \mathrm{M}_\odot$). For quantities subject to change because of precession, we quote values at a reference gravitation-wave frequency of $f_{\mathrm{ref}} = 20$ Hz. There are small differences in the source's parameters compared to the runs on the older calibration, but these are well within the total uncertainty; the greatest difference is in the sky area, where the reduced calibration uncertainty improves the localization area by a factor of about 2–3.

There are two differences in the configuration of the analysis of LVT151012 from that for GW150914: The prior on the component masses was set to be uniform over the range $m_{1,2} \in [5, 80] \mathrm{M}_\odot$, and the length of data analyzed was $T = 22$ s. We find that LVT151012 is consistent with a lower-mass source, which necessitates a lower prior bound on the component masses and requires us to analyze a longer stretch since the signal could be in band for longer.

GW151226 is also consistent with being a lower-mass source. However, we can still consider just 8 s of data by confining the component masses such that the chirp mass is $\mathcal{M} \in [9.5, 10.5] \mathrm{M}_\odot$ and the mass ratio is $q \in [1/18, 1]$. Preliminary analyses found no support outside of these ranges, and the final posteriors lie safely within this region. This choice of segment length limits the computational expense of the analysis.

## APPENDIX C: RATES CALCULATION DESCRIPTION

In this appendix, we give further details of how the BBH coalescence rates are estimated. The framework of Ref. [229] considers two classes of triggers (coincident search events): those whose origin is astrophysical and those whose origin is terrestrial. Terrestrial triggers are the result of either instrumental or environmental effects in the detector. In order to calculate the rate of astrophysical triggers, we first seek to determine the probability that any given trigger arises from either class. The two classes of source produce triggers with different densities as a function of the detection statistic used in the analysis, which we denote as $x$. Triggers appear in a Poisson process with number density

$$\frac{\mathrm{d}N}{\mathrm{d}x} = \Lambda_1 p_1(x) + \Lambda_0 p_0(x), \tag{C1}$$

where $\Lambda_1$ and $\Lambda_0$ are the Poisson mean numbers of triggers of astrophysical and terrestrial origin, respectively. Here, $\Lambda_1$ is related to the merger rate density through

$$\Lambda_1 = R \langle VT \rangle, \tag{C2}$$

where $\langle VT \rangle$ is the population-averaged sensitive space-time volume of the search [42,155],

$$\langle VT \rangle = T \int \mathrm{d}z \mathrm{d}\theta \frac{\mathrm{d}V_c}{\mathrm{d}z} \frac{1}{1+z} s(\theta) f(z, \theta), \tag{C3}$$

where $V_c$ is the comoving volume [230], $\theta$ describes the population parameters, $s(\theta)$ is the distribution function for the astrophysical population in question, and $0 \leq f(z, \theta) \leq 1$ is the selection function giving the probability of detecting a source with parameters $\theta$ at redshift $z$. Because the distribution of astrophysical triggers is independent of source parameters without parameter-estimation follow-up, we must *assume* an astrophysical distribution of sources, and the rate enters the likelihood only in the form $\Lambda_1 = R\langle VT \rangle$. We also marginalize over a calibration uncertainty of 6% on the recovered luminosity distances (18% uncertainty on $\langle VT \rangle$) when computing the rates.

The distribution of terrestrial triggers is calculated from the search background estimated by the analyses (as shown in Fig. 3). The distribution of astrophysical events is determined by performing large-scale simulations of signals drawn from the various astrophysical populations added to the O1 data set and using the distribution of triggers recovered by our detection analyses applied to this data set. This method correctly accounts for various thresholds applied in the analyses. Note that the observed distribution of astrophysical triggers over the detection statistic will be essentially independent of the astrophysical population used: All populations are assumed to be distributed uniformly in comoving volume; thus, to a good approximation, the measured SNRs and other detection statistics follow the flat-space, volumetric density $p_1(\rho) \propto \rho^{-4}$ [129].

The likelihood for a search result containing $M$ triggers with detection statistic values $\{x_j | j = 1, \ldots, M\}$ is [229]

$$\mathcal{L}(\{x_j|j=1,\ldots,M\}|\Lambda_1,\Lambda_0) = \left\{\prod_{j=1}^{M}[\Lambda_1 p_1(x_j)+\Lambda_0 p_0(x_j)]\right\}\exp[-\Lambda_1-\Lambda_0]. \quad (C4)$$

The posterior over $\Lambda_1$ and $\Lambda_0$ is then obtained by multiplying the likelihood in Eq. (C4) by a prior proportional to $1/\sqrt{\Lambda_0\Lambda_1}$ and marginalizing over the $x_j$ to obtain $p(\Lambda_0,\Lambda_1)$. For a trigger with statistic value $x$, the probability that it is of astrophysical origin is

$$P_1(x|\{x_j|j=1,\ldots,M\}) \equiv \int d\Lambda_0 d\Lambda_1 \frac{\Lambda_1 p_1(x)}{\Lambda_0 p_0(x)+\Lambda_1 p_1(x)} \times p(\Lambda_1,\Lambda_0|\{x_j|j=1,\ldots,M\}). \quad (C5)$$

Finally, we evaluate the rate assuming a population containing only BBH mergers with mass and spin parameters matching the three triggers for which $P_1 > 0.5$; i.e., astrophysical origin is more likely than terrestrial. To do so, we must generalize the formalism presented above to account for three different astrophysical populations, each having a different mean number of triggers $\Lambda_i$. In this case, the likelihood of Eq. (C4) is generalized to allow for each trigger to arise from one of the astrophysical classes, or be of terrestrial origin. Additionally, we change the prior distribution to account for the number of astrophysical trigger classes via

$$p(\{\Lambda_i\},\Lambda_0) \propto \left(\sum_i^{N_c}\Lambda_i\right)^{-N_c+1/2}\Lambda_0^{-1/2}, \quad (C6)$$

where $N_c = 3$ is the number of different classes of astrophysical triggers. This functional form is chosen to prevent the posterior expectation of the total count of astrophysical events, $\sum_i^{N_c}\Lambda_i$, from growing without limit as more classes are considered in the calculation.

The three triggers associated with GW150914, GW151226, and LVT151012 are restricted to originate either from their specific class or be of terrestrial origin. Thus, for instance, we neglect any probability of GW150914 arising from the class containing GW151226. We justify this by noting that the probability distributions for the component masses of the three likely signals are disjoint from one another at high confidence.

Multiplying this prior by the generalization of the likelihood, Eq. (C4), we obtain the posterior distribution on $\Lambda_i$, the number of astrophysical triggers in each class. We again calculate the sensitive $\langle VT\rangle$ for each of the classes of signals and thus infer merger rates for each class. Figure 14 shows how the sensitive $\langle VT\rangle$ is accumulated as a function of redshift. For the less massive GW151226, the peak occurs at $z \sim 0.1$, while for GW150914, it occurs at $z \approx 0.2$, with the search being sensitive to some signals with redshifts as high as 0.6.

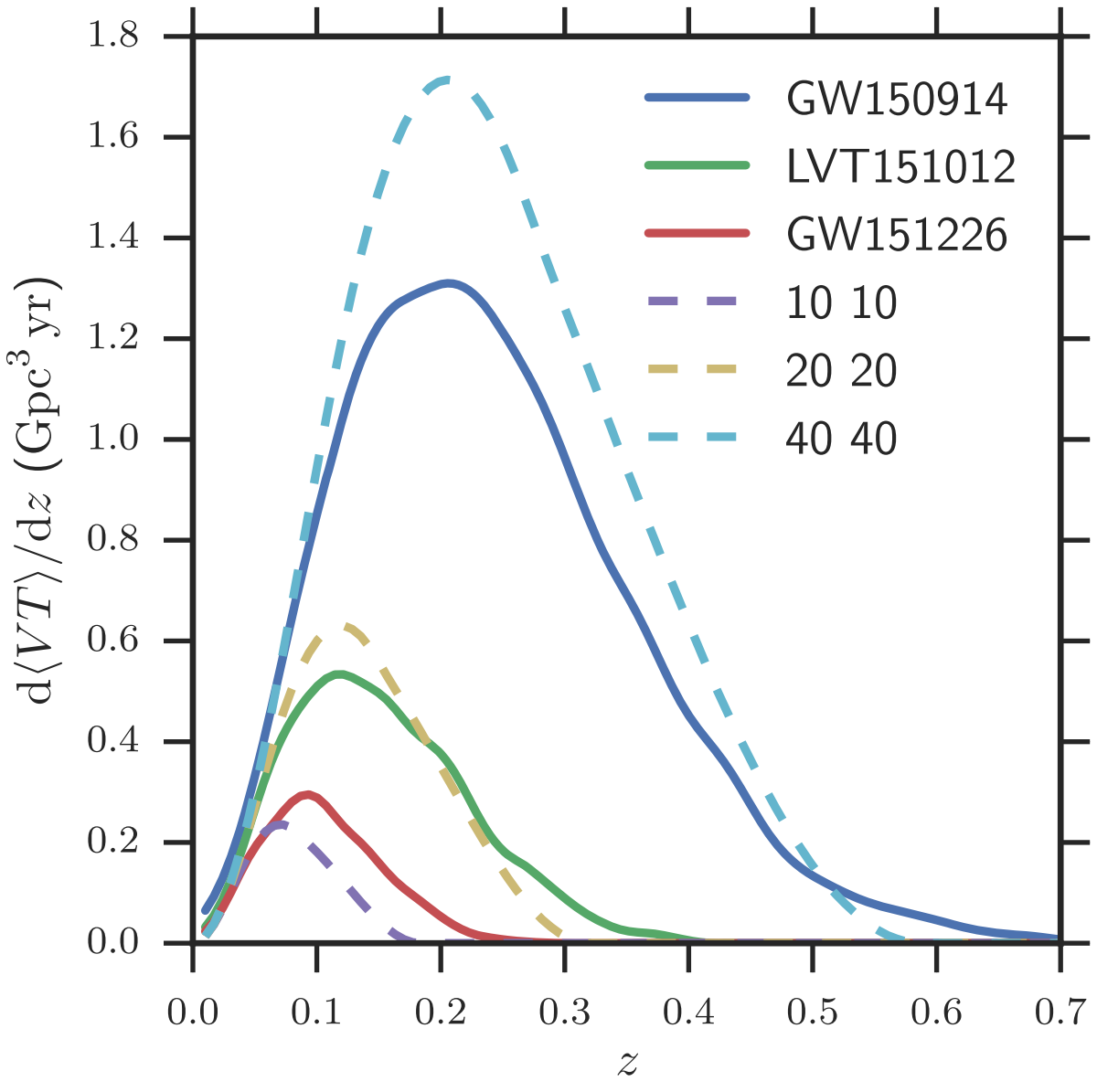

FIG. 14. The rate at which sensitive space-time volume accumulates with redshift. Curves labeled by component masses in $M_\odot$ are computed using an approximate prescription described in Ref. [42], assuming sources with fixed masses in the comoving frame and with zero component spins; the GW150914, GW151226, and LVT151012 curves are determined from the Monte Carlo injection campaign described in Sec. VI.

## APPENDIX D: MASS DISTRIBUTION CALCULATION DESCRIPTION

Here, we describe the details of the analysis of the mass distribution that appears in Sec. VI. Further details on population analysis in the context of measurement uncertainty and selection effects are given in Ref. [231]. After this paper was accepted, we became aware of Ref. [232], which derives the same result as Ref. [231] and this appendix in a different context. Useful references for hierarchical analysis in astronomy include Refs. [156–159].

We assume that the distribution of black hole masses in coalescing binaries follows [see Eq. (7)]

$$p(m_1) \propto m_1^{-\alpha} \quad (D1)$$

and

$$p(m_2|m_1) = \frac{1}{m_1 - M_{\min}}, \quad (D2)$$

with $M_{\min} = 5M_\odot$ the minimum black hole mass we consider, as in the models of the mass distributions used to infer rates. The joint population distribution on $m_1$ and $m_2$ therefore follows

$$p(m_1, m_2|\alpha) \propto \frac{m_1^{-\alpha}}{m_1 - M_{\min}}. \tag{D3}$$

Here, we take all masses to be source-frame masses. The distribution of masses observed in our experiment will differ from the population distribution because our detector sensitivity is a strong function of system mass.

A simplified model of our detection pipeline is that it is a deterministic function of the data, $f(d)$, such that when $f(d) > f_0$, for some threshold $f_0$, we detect a trigger. Given our population parameter $\alpha$, the joint distribution of system parameters and data for a single *detected* trigger with data $d$ is

$$p(d, m_1, m_2|\alpha) = \frac{p(d|m_1, m_2)p(m_1, m_2|\alpha)}{\beta(\alpha)}, \tag{D4}$$

where the first term in the numerator is the standard (unnormalized) likelihood function used in our parameter-estimation analysis, the second term is the population distribution in Eq. (D3) and plays a role of a prior in our hierarchical analysis, and $\beta(\alpha)$ is a normalization factor, ensuring that the joint distribution is properly normalized. This factor is

$$\beta(\alpha) = \int \mathrm{d}m_1 \mathrm{d}m_2 \mathrm{d}d\, p(d|m_1, m_2)p(m_1, m_2|\alpha), \tag{D5}$$

where the integral is taken over all allowed masses and the set of data producing a detected trigger $\{d|f(d) > f_0\}$.

Consider first the integral over $d$, which is

$$\int_{\{d|f(d)>f_0\}} \mathrm{d}d\, p(d|m_1, m_2)p(m_1, m_2|\alpha) = p(m_1, m_2|\alpha)P_{\mathrm{det}}(m_1, m_2), \tag{D6}$$

where we have defined the detection probability as a function of mass,

$$P_{\mathrm{det}}(m_1, m_2) \equiv \int_{\{d|f(d)>f_0\}} \mathrm{d}d\, p(d|m_1, m_2). \tag{D7}$$

This quantity is proportional to the $\langle VT \rangle$ defined in Eq. (C3) evaluated with a source distribution that fixes the source masses:

$$P_{\mathrm{det}}(m_1, m_2) \propto \langle VT \rangle|_{m_1, m_2}. \tag{D8}$$

To evaluate this factor, we use the approximate recipe from Ref. [42]. Thus,

$$\beta(\alpha) \propto \int \mathrm{d}m_1 \mathrm{d}m_2 p(m_1, m_2|\alpha)\langle VT \rangle|_{m_1, m_2}. \tag{D9}$$

This normalization factor accounts for the selection effects of our searches on the observed distribution of masses.

Here, we are interested only in the population parameters, not in reanalyzing the system masses; thus, we can integrate the masses out of the joint distribution in Eq. (D4) to obtain

$$p(d|\alpha) = \frac{1}{\beta(\alpha)} \int \mathrm{d}m_1 \mathrm{d}m_2 p(d|m_1, m_2)p(m_1, m_2|\alpha) \propto \frac{1}{\beta(\alpha)} \langle p(m_1, m_2|\alpha) \rangle, \tag{D10}$$

where the notation $\langle \ldots \rangle$ refers to an average over posterior samples (properly reweighted to correspond to a flat prior in $m_1$ and $m_2$) [231].

With multiple triggers analyzed, the likelihood is a product of single-event likelihoods from Eq. (D10). We impose a flat prior on $\alpha$. The posterior from an analysis using GW150914, LVT151012, and GW151226 appears in Fig. 12.

# Erratum: Binary Black Hole Mergers in the first Advanced LIGO Observing Run [Phys. Rev. X 6, 041015 (2016)]

LIGO Scientific and Virgo Collaborations*

This *Erratum* reports an error found in the implementation of the code of the LIGO Scientific and Virgo Collaborations (LVC) as used in gravitational-wave based estimations of possible deviations from the post-Newtonian (PN) terms expected in general relativity. The error concerned the 0.5 PN term, and affected the results previously published for GW150914 [1] in **PRL 116.221101 (2016)** [2], for GW151226 [3] in **PRX 6.041015 (2016)** [4], and for GW170104 in **PRL 118.221101 (2017)** [5]. We corrected the bug, and present the reproduced results in this *Erratum* , as well as in the respective Errata papers [6, 7]. The main conclusion, that the results are consistent with general relativity, remains.

The test for the parameterized post-Newtonian [8] deviations from the expected GR values, relied on creating non-GR waveforms [2, 9–13], and using them as potential matches for the observed waveforms [14–17]. In these waveforms, implemented in the frequency domain, freedom was introduced by allowing the phase coefficients describing different powers of the post-Newtonian parameter (equivalently, powers of the frequency) to assume a range of values, not only the particular values prescribed by GR.

However, a coding bug was introduced, identically zeroing the deviations at 0.5PN in the inspiral regime (as in GR). The 0.5PN deviations were hence absent in the phasing formula, though not in the junction conditions that relate the inspiral regime to the intermediate regime. Any constraints obtained in [2, 4, 5] only resulted from the latter.

This error affected the results of the non-GR parameter estimation (PE) [14] pipeline tests performed for finding bounds on possible PN deviations from GR. In particular, they affect the bounds on the single deviations in the 0.5 PN term, and on the tests with multiple deviations together. These erroneous results appeared in Figs. 6 & 7 and Table I of [2], in Figs. 7 & 8 of [4], and in Fig. 9 of the Supplemental Material of [5]. The corrected versions of all of these have been produced. The corrections for 7 & 8 of [4] appear below, while the others are available in [6, 7]. All these results remain consistent with GR.

The error, introduced by erroneous caching during the optimization of the waveform generation for efficient PE, has been corrected in commit 980b3f788a445d135390eba8861b03f70133cd47 of the LALSuite[18] code. No subsequent LVC papers have been affected.

**Note**: While this error also affected the analysis of GW170608 [19], the reported results require no changes: with the corrected analysis, the GR predicted PN coefficient values continue to be consistent with the data. No change is required regarding the preliminary reported results for GW170814 [20], either.

**Note**: We also correct the value of $\ln\mathcal{L}$ for GW151226, given in Section III B: while previously reported as 22.6, which was the value from the original pipeline, the final correct value for the GstLAL pipeline is 33.7.

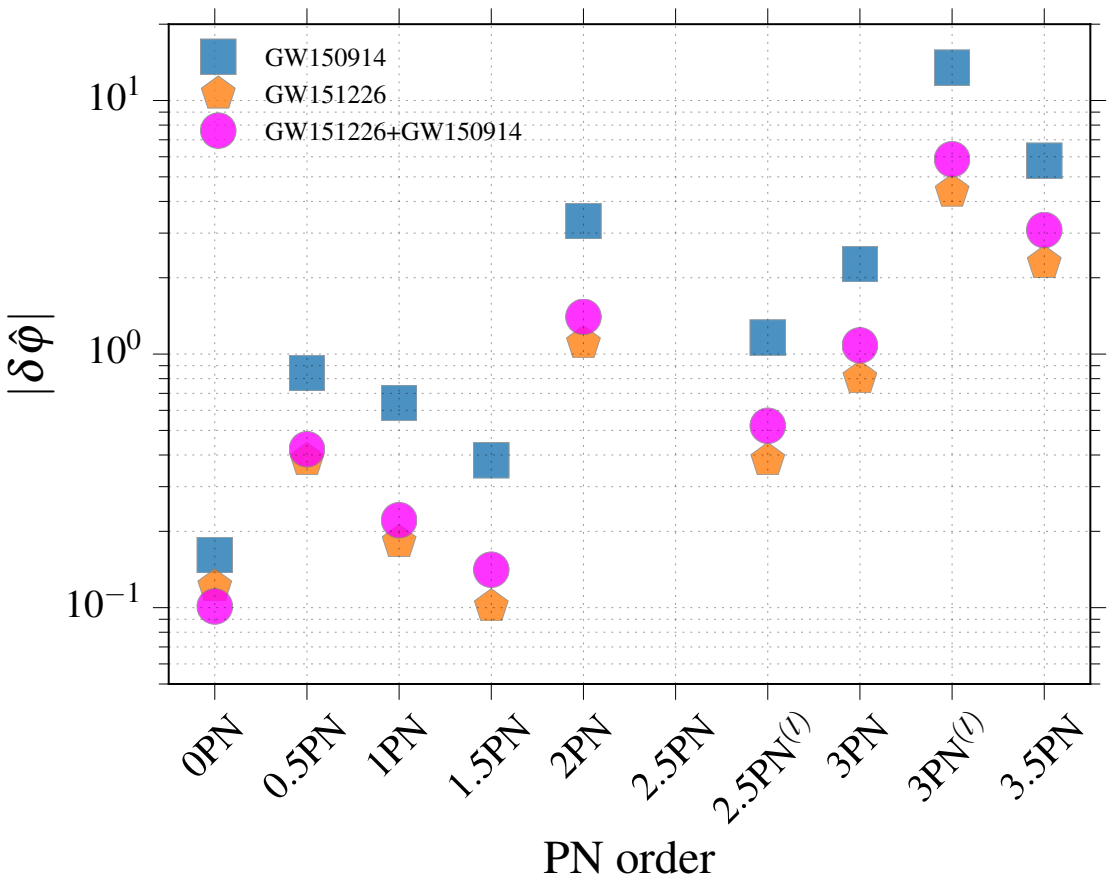

FIG. 1: This is the corrected Fig. 8. of [4], showing the 90% credible upper bounds on deviations in the PN coefficients from GW150914, from GW151226 and from both. It also fixes a previous plotting error for the 0PN individual bounds (the GW150914 and GW151226 bounds were incorrectly drawn on top of eachother).

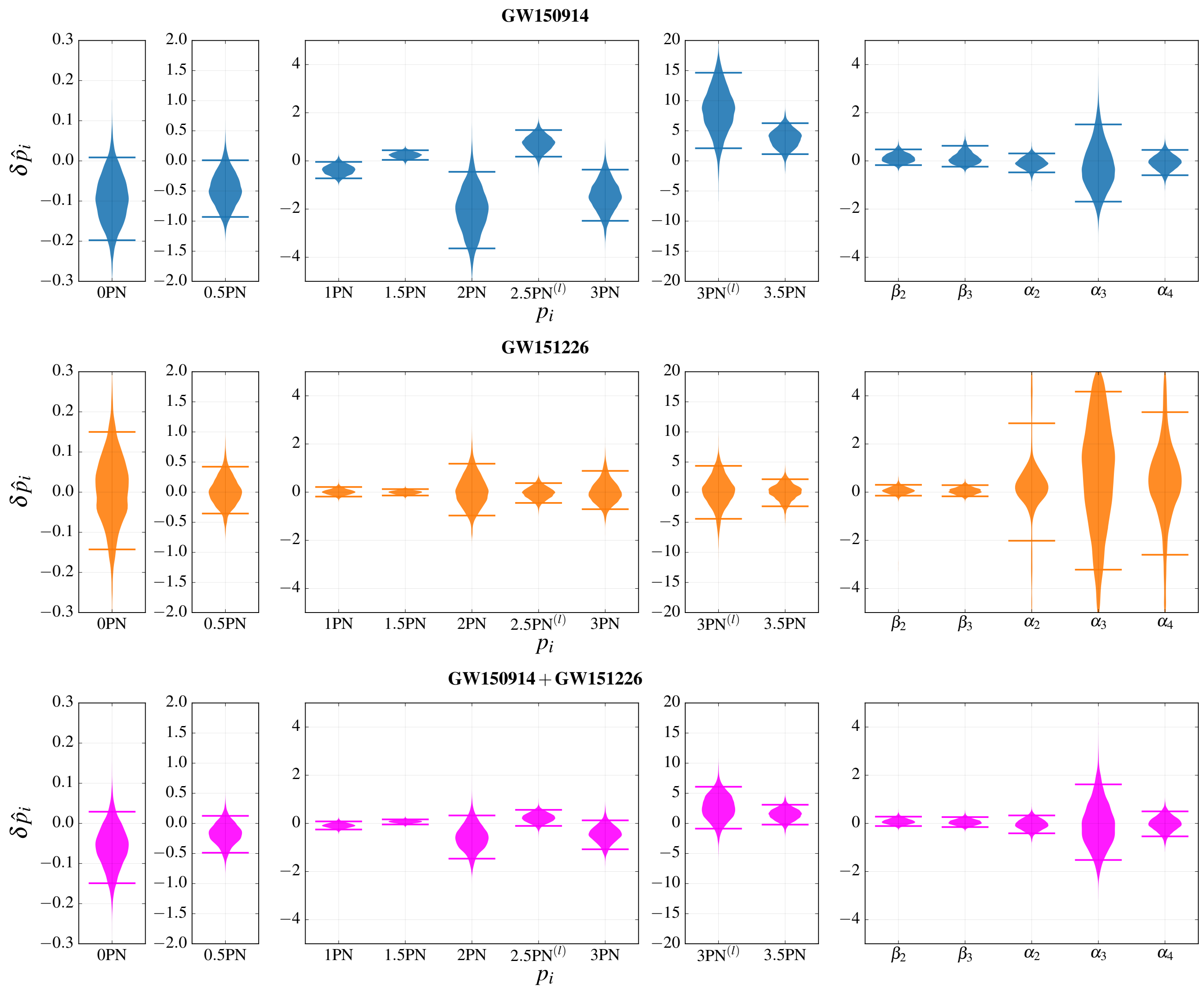

.

FIG. 2: This is the corrected Fig. 7. of [4], showing the posterior density distributions and 90% credible intervals for the relative deviations PN parameters $p_i$, as well as intermediate parameters $\beta_i$ and merger-ringdown parameters $\alpha_i$. The corrected posterior for the 0.5PN single-parameter analysis peaks below 0, while previously reported to peak above; the posterior is still consistent with GR. This plot incorporates the results of [21].

* lsc-spokesperson@ligo.org; Virgo-spokesperson@ego-gw.it